\pdfoutput=1
\documentclass[12pt]{clear2027}
\jmlrproceedings{}{Preprint}\jmlrvolume{}\jmlrpages{}\jmlryear{}

\usepackage{times}
\usepackage[utf8]{inputenc} 
\usepackage[T1]{fontenc}    
\usepackage{microtype}
\usepackage{tikz}
\usetikzlibrary{positioning,arrows.meta,fit,calc}
\usepackage{booktabs} 
\usepackage{makecell}
\usepackage{adjustbox}

\definecolor{linkcolor}{RGB}{83,83,182}

\hypersetup
{
  colorlinks=true,
  citecolor=linkcolor,
  linkcolor= linkcolor,
  urlcolor=blue,
  pdfstartview=FitH,
  pdftitle={},
  pdfauthor={},
  pdfcreator={},
  pdfsubject={},
  pdfkeywords={},
}

\usepackage{mathtools}
\usepackage{thmtools}
\usepackage{thm-restate}

\usepackage[capitalize,noabbrev]{cleveref}

\newcommand{\E}{\mathbb E}
\newcommand{\1}{\mathbf 1}
\newcommand{\PXT}{P_{XT}}

\graphicspath{
  {figures/}
  {../figures/}
}

\crefname{table}{Tab.}{Tabs.}
\crefname{figure}{Fig.}{Figs.}
\crefname{section}{Sec.}{Secs.}
\crefname{theorem}{Thm}{Thms}
\crefname{proposition}{Prop.}{Props.}
\crefname{corollary}{Cor.}{Cors.}
\crefname{lemma}{Lem.}{Lems.}
\crefname{appendix}{App.}{Apps.}

\title[Balancing Partition Forest]{Balancing Partition Forest for Causal Weighting}

\clearauthor{%
 \Name{Felipe Angelim} \Email{felipeangelim@pm.me}\\
 \addr Independent Researcher
 \AND
 \Name{Alessandro Leite} \Email{aleite@insa-rouen.fr}\\
 \addr INSA Rouen Normandy, Normandy University, LITIS, Rouen, France%
}

\begin{document}
\addtocontents{toc}{\protect\setcounter{tocdepth}{-10}}

\maketitle

\begin{abstract}
  Balancing weights are used in observational causal inference to construct a
  population in which the treatment is independent of observed pretreatment
  covariates. The ideal weight is the ratio between the product of the covariate
  and treatment marginals and their joint distribution, for binary and
  continuous treatments alike. We propose Balancing Partition Tree (BPT) and
  Balancing Partition Forest (BPF), which estimate this ratio on a learned
  rectangular partition of covariate--treatment space. Each split maximizes the
  gain in a Bregman divergence computed from joint and marginal counts. Two
  choices of divergence yield criteria that maximize the mutual information
  captured by the partition or reduce the squared error of the weights. We show
  that the population residual of either criterion bounds the bias of
  inverse-probability-weighted and doubly robust response estimators that use
  the population partition weights. Across 17 synthetic and semi-synthetic
  settings and three response estimators, BPF obtains the lowest mean RMSE rank
  among seven weight estimators, and a single tree fits in 0.14 seconds on
  16,512 observations.%
\end{abstract}


\section{Introduction}\label{sec:introduction}

Estimating causal effects from observational data requires accounting for systematic differences in treatment assignment across individuals. When treatment depends on pretreatment covariates, comparisons across treatment levels may conflate treatment effects with differences in the underlying covariate distributions. A widely used strategy is to reweight the observed data to construct a population in which treatment and covariates are independent \citep{rosenbaum1983central, robins2000marginal}. Let \(P_{XT}\) denote the observed joint distribution of covariates \(X\) and treatment \(T\), with marginals \(P_X\) and \(P_T\). The target distribution is their product \(Q=P_X\otimes P_T\), and the corresponding stabilized balancing weight is the density ratio \(w=dQ/dP_{XT}\). This formulation applies to binary, categorical, and continuous treatments, providing a unified view of balancing-weight estimation.

Despite this simple characterization of the population, estimating balancing weights remains challenging. Propensity-score and conditional-density approaches require modeling treatment assignment, potentially producing inaccurate or highly variable weights when treatment--covariate relationships are complex. Covariate-balancing methods instead target selected balance conditions, which need not eliminate general forms of dependence. More flexible density-ratio approaches, such as permutation weighting~\citep{arbour2021permutation}, learn to distinguish observed treatment--covariate pairs from pairs generated under an empirical independence distribution. Although this avoids specifying a conditional treatment model, generating reference pairs and fitting classifiers can introduce additional computational costs, particularly when multiple permutations are used. These considerations motivate an alternative research question which comprises: \textbf{can balancing weights be learned flexibly and efficiently by directly exploiting the structure of the independence distribution, without explicitly generating reference observations?}

We address this question by proposing \textbf{Balancing Partition Tree (BPT)} and \textbf{Balancing Partition Forest (BPF)}, which are adaptations of Partition Tree and Partition Forest~\citep{angelim2026partition} for causal weighting. Our central observation is that rectangular partitions of covariate--treatment space provide a natural representation of the target density ratio. For a cell \(A=A_X\times A_T\), the product structure of \(Q\) gives \(Q(A)=P_X(A_X)P_T(A_T)\). Consequently, its balancing weight can be estimated directly as the ratio of the product of empirical marginal masses to the empirical joint mass. This reduces cell-wise density-ratio estimation to counting observations, without fitting a conditional treatment density or generating permuted samples.

\begin{figure}[t]
  \centering
\input{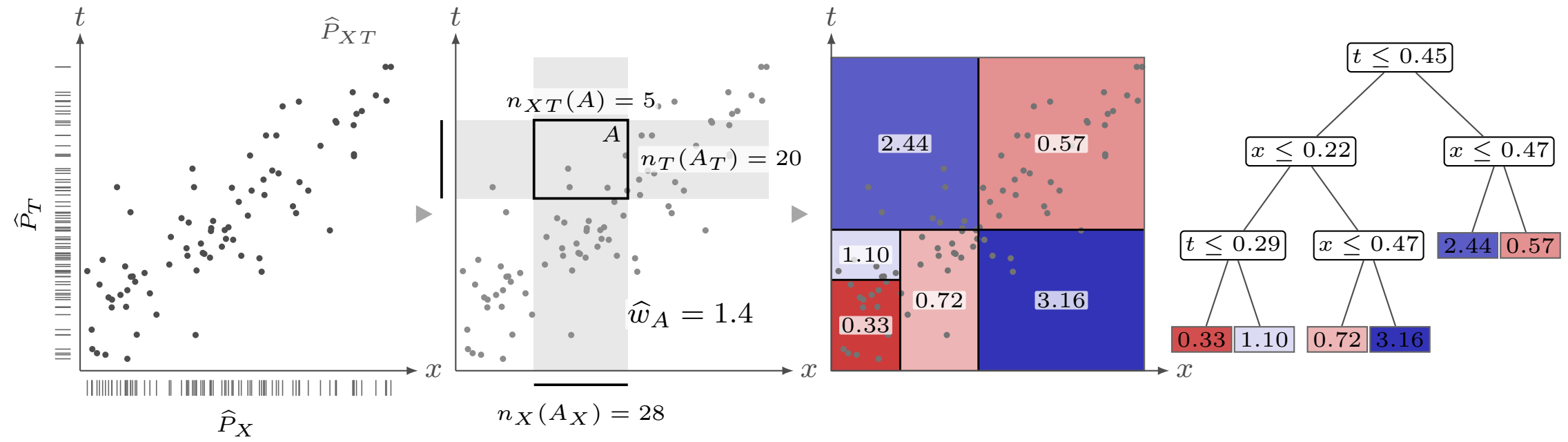}
\colorlet{wred}{red!75!black}
\colorlet{wblue}{blue!65!black}
\begin{tikzpicture}[
  font=\scriptsize,
  pt/.style={circle, fill=black!70, inner sep=0pt, minimum size=1.7pt},
  ax/.style={-{Latex[length=1.1mm]}, black!70, line width=0.4pt},
  rug/.style={black!60, line width=0.3pt},
  split/.style={black, line width=0.55pt},
  tn/.style={draw, rounded corners=1pt, fill=white, inner sep=1.2pt, font=\tiny},
  leafn/.style={draw=black!60, inner sep=1.5pt, font=\tiny},
  edge/.style={black!70, line width=0.4pt},
  wlab/.style={font=\tiny, fill=white, fill opacity=0.75, text opacity=1, inner sep=0.6pt, rounded corners=0.5pt},
  title/.style={anchor=north, align=center, font=\footnotesize},
  flow/.style={-{Triangle[length=1.6mm, width=1.6mm]}, black!35, line width=1pt},
]
\def\S{3.0cm}
\def\gap{3.6cm}
\def\titleY{-0.68cm}

\begin{scope}[x=\S, y=\S]
  \draw[ax] (0,0) -- (1.08,0) node[right=-2pt] {$x$};
  \draw[ax] (0,0) -- (0,1.08) node[above=-2pt] {$t$};
  \foreach \x/\t in \pts {
    \node[pt] at (\x,\t) {};
    \draw[rug] (\x,-0.03) -- (\x,-0.08);
    \draw[rug] (-0.03,\t) -- (-0.08,\t);
  }
  \node[font=\tiny, anchor=north] at (0.5,-0.08) {$\widehat P_X$};
  \node[font=\tiny, anchor=south, rotate=90] at (-0.08,0.5) {$\widehat P_T$};
  \node[font=\tiny, anchor=south east, text=black!70] at (1,1.0) {$\widehat P_{XT}$};
\end{scope}
\node[title] at (0.5*\S,\titleY) {(a) Observed pairs};

\begin{scope}[x=\S, y=\S, shift={(\gap,0)}]
  \def\xa{0.25}\def\xb{0.55}\def\ta{0.55}\def\tb{0.80}
  \fill[black!9] (\xa,0) rectangle (\xb,1);
  \fill[black!9] (0,\ta) rectangle (1,\tb);
  \draw[ax] (0,0) -- (1.08,0) node[right=-2pt] {$x$};
  \draw[ax] (0,0) -- (0,1.08) node[above=-2pt] {$t$};
  \foreach \x/\t in \pts { \node[pt, fill=black!45] at (\x,\t) {}; }
  \draw[black, line width=0.7pt] (\xa,\ta) rectangle (\xb,\tb);
  \node[wlab] at ({(\xa+\xb)/2},\tb+0.06) {$n_{XT}(A)=5$};
  \node[wlab, anchor=west] at (\xb+0.03,{(\ta+\tb)/2}) {$n_T(A_T)=20$};
  \node[font=\tiny, anchor=north east, inner sep=1.5pt] at (\xb,\tb) {$A$};
  \node[wlab, font=\scriptsize] at (0.75,0.18) {$\widehat{w}_A=1.4$};
  \draw[black, line width=0.7pt] (\xa,-0.045) -- (\xb,-0.045)
    node[midway, below=1pt, font=\tiny] {$n_X(A_X)=28$};
  \draw[black, line width=0.7pt] (-0.045,\ta) -- (-0.045,\tb);
\end{scope}
\node[title] at (\gap+0.5*\S,\titleY) {(b) Cell weight};

\begin{scope}[x=\S, y=\S, shift={(2*\gap,0)}]
  \foreach \xa/\xb/\ta/\tb/\w/\c in \leaves { \fill[\c] (\xa,\ta) rectangle (\xb,\tb); }
  \foreach \x/\t in \pts { \node[pt, fill=black!55] at (\x,\t) {}; }
  \draw[split] (0,0.45) -- (1,0.45);
  \draw[split] (0.22,0) -- (0.22,0.45);
  \draw[split] (0,0.29) -- (0.22,0.29);
  \draw[split] (0.47,0) -- (0.47,1);
  \draw[black!60, line width=0.4pt] (0,0) rectangle (1,1);
  \foreach \xa/\xb/\ta/\tb/\w/\c in \leaves {
    \node[wlab] at ({(\xa+\xb)/2},{(\ta+\tb)/2}) {$\w$};
  }
  \draw[ax] (0,0) -- (1.08,0) node[right=-2pt] {$x$};
  \draw[ax] (0,0) -- (0,1.08) node[above=-2pt] {$t$};
\end{scope}
\begin{scope}[shift={(2*\gap+\S+0.55cm,3.0cm)}, y=-1cm, x=1cm]
  \node[tn] (r)  at (1.9,0)    {$t\le0.45$};
  \node[tn] (l)  at (0.95,0.9) {$x\le0.22$};
  \node[tn]         (rr) at (2.85,0.9) {$x\le0.47$};
  \node[tn]         (ll) at (0.3,1.8)  {$t\le0.29$};
  \node[tn]         (lr) at (1.6,1.8)  {$x\le0.47$};
  \node[leafn, fill=wblue!64!white] (rrl) at (2.55,1.8) {$2.44$};
  \node[leafn, fill=wred!43!white]  (rrr) at (3.15,1.8) {$0.57$};
  \node[leafn, fill=wred!69!white]  (lll) at (0,2.7)    {$0.33$};
  \node[leafn, fill=wblue!14!white] (llr) at (0.6,2.7)  {$1.10$};
  \node[leafn, fill=wred!29!white]  (lrl) at (1.3,2.7)  {$0.72$};
  \node[leafn, fill=wblue!80!white] (lrr) at (1.9,2.7)  {$3.16$};
  \foreach \p/\c in {r/l, r/rr, l/ll, l/lr, rr/rrl, rr/rrr, ll/lll, ll/llr, lr/lrl, lr/lrr}
    \draw[edge] (\p) -- (\c);
\end{scope}
\node[title] at (2*\gap+0.5*\S,\titleY) {(c) Partition and tree};

\foreach \i in {0,1} {
  \draw[flow] (\i*\gap+\S+0.22cm,0.5*\S) -- (\i*\gap+\gap-0.22cm,0.5*\S);
}
\end{tikzpicture}
  \caption{Balancing partition trees on 80 observed pairs.
  (a)~Treatment--covariate dependence.
  (b)~Each cell's weight is computed from its joint and marginal counts.
  (c)~The partition and its corresponding tree, with leaf weights shown
  numerically. Red cells are down-weighted; blue cells are up-weighted
  to approximate independence.}
  \label{fig:method-overview}
\end{figure}

The partition is learned adaptively by selecting splits that improve a Bregman-divergence objective. We develop two instances of this framework: a mutual-information criterion that greedily increases the treatment--covariate dependence captured by the partition, and a squared-error criterion that reduces the population approximation error of the balancing weights. Both criteria depend only on joint and marginal cell counts and admit efficient split evaluation through cumulative statistics. Because marginal splits alone cannot reveal dependence at the root, the tree also permits exploration splits that expose treatment--covariate interactions. To reduce sensitivity to individual partitions, BPF aggregates bootstrap trees through the harmonic mean of their estimated weights. \Cref{fig:method-overview} illustrates the resulting construction.

Beyond the computational formulation, we establish a theoretical connection between partition quality and causal estimation error. A Bregman decomposition separates the approximation error induced by the partition from the error associated with estimating its cell weights. For the two splitting objectives, the population approximation residual can be expressed, respectively, in terms of residual mutual information and squared weight error. Under appropriate regularity conditions, these quantities yield bounds on the bias of inverse-probability-weighted and doubly robust causal-response estimators when population partition weights are used. Thus, the partition criteria are connected not only to density-ratio approximation but also to the downstream causal estimation problem.

We evaluate the proposed methods across seven synthetic data-generating processes, two sample sizes, and three semi-synthetic datasets, yielding 17 experimental settings. We compare seven weight estimators for response estimation using the Marginal Structural Model~\citep{robins2000marginal}, the Generalized Propensity Score~\citep{hirano2004propensity}, and the Doubly Robust~\citep{kennedy2017nonparametric} approaches under a common tuning protocol. BPF achieves the lowest overall mean response-RMSE rank under all three estimators, with ranks of 2.29, 2.12, and 2.24, respectively. Forest aggregation generally improves accuracy over individual partition trees, while weight-range diagnostics yield fewer extreme estimates than some competing approaches across the evaluated settings. The method also offers efficient fitting: on 16,512 observations, a single BPT fits in 0.142 seconds, compared with 107.478 seconds for the evaluated permutation-weighting implementation using 100 permutations, under the single-thread, fixed-configuration timing protocol.

\textbf{Our contributions are threefold}. \textbf{First}, we introduce an adaptive partition-based framework for balancing-weight estimation, with two count-based Bregman splitting criteria and harmonic forest aggregation (\cref{sec:methodology}). \textbf{Second}, we establish a Bregman risk decomposition and population causal-bias bounds that connect the quality of partition approximations to weighted and doubly robust estimation (\cref{sec:properties}). \textbf{Third}, we provide an empirical evaluation across binary and continuous treatments, multiple causal-response estimators, and diverse synthetic and semi-synthetic settings, including analyses of weight stability and computational cost (\cref{sec:experiments}).

\section{Related work}\label{sec:related-work}

\paragraph{Balancing weights.}
Inverse-probability weighting has roots in the Horvitz--Thompson estimator
\citep{horvitz1952generalization}. \citet{rosenbaum1983central} established the
balancing property of the propensity score, and \citet{robins2000marginal}
used inverse-probability weights to fit marginal structural models.
Later methods target sample balance directly. Entropy balancing
\citep{hainmueller2012entropy} calibrates weights to covariate moments, CBPS
\citep{imai2014covariate,fong2018covariate} adds balance conditions to the propensity
model for binary and continuous treatments, and stable balancing weights
\citep{zubizarreta2015stable} minimize weight variance under approximate
balance. For continuous treatments, independence weights
\citep{huling2024independence} minimize a measure of dependence between
treatment and covariates.

\paragraph{Density ratios and permutation weighting.}
Permutation weighting \citep{arbour2021permutation} estimates balancing weights
with a classifier that separates observed pairs from pairs with permuted
treatments, and relates classifier regret to causal estimation error through
Bregman divergences. Averaging over permutations reduces their variability
at the cost of additional classifier fits. We obtain the reference mass of
each cell directly from marginal counts, so fitting a tree requires no
permuted pairs. Our tuning protocol in the experiments section uses the held-out
permutation log loss of \citet{arbour2021permutation}.

\paragraph{Density-ratio estimation.}
Direct density-ratio estimators fit the ratio without estimating the two
densities, by Kullback--Leibler importance estimation
\citep{sugiyama2007direct}, by least squares \citep{kanamori2009least} or by
convex risk minimization \citep{nguyen2010estimating}.
\citet{sugiyama2012density} unify these methods as density-ratio matching
under a Bregman divergence. Our two criteria belong to this family, restricted
to piecewise-constant functions on a learned partition and with the product
of the empirical marginals as the reference sample.

\paragraph{Adaptive partitions and information divergence.}
\citet{darbellay1999estimation} estimate mutual information on adaptive
rectangular partitions refined at marginal quantiles. On a fixed partition,
our cell weights are the reciprocals of their joint-to-product ratios.
Their decomposition of mutual information into a partition term and a
within-cell residual is a precursor of our residual mutual information.
\citet{silva2010nonproduct} give consistency conditions for data-dependent,
non-product partitions, and \citet{wang2005divergence} estimate divergences
on data-dependent partitions. We select splits by a Bregman gain
\citep{banerjee2005clustering} and relate the remaining residual to causal
bias in \cref{sec:properties}.

\section{Methodology}\label{sec:methodology}

\subsection{Notation and setup}

We observe a dataset $\mathcal D=\{(x_i,t_i,y_i)\}_{i=1}^N$
of independent observations drawn from the distribution of
$(X,T,Y)$, where $X\in\mathcal X$ denotes pretreatment covariates,
$T\in\mathcal T$ denotes treatment, and $Y\in\mathbb R$ denotes
the outcome. Both $\mathcal X$ and $\mathcal T$ may contain
continuous and categorical coordinates. Let $Y(t)$ denote the
potential outcome under treatment $t$. Our downstream target is
the Average Response Function (ARF) $\mu(t)=\mathbb E[Y(t)]$, assumed finite.

Balancing weights aim to make treatment independent of the
observed covariates while preserving their marginal distributions.
Let $P_{XT}$ denote the joint distribution of $(X,T)$, with
marginals $P_X$ and $P_T$. The target distribution is therefore
\[
    Q := P_X\otimes P_T.
\]
We assume that $P_{XT}$ and $Q$ are mutually absolutely continuous.
This allows us to define the stabilized weight:
\begin{equation}
    w := \frac{dQ}{dP_{XT}}.
\end{equation}
Since
$w\,dP_{XT}=dQ$, weighting the joint distribution by $w$
makes $X$ and $T$ independent.

A causal interpretation additionally requires consistency,
$Y=Y(T)$, and conditional exchangeability,
$Y(t)\perp T\mid X$ for each treatment value under consideration.
The condition $Q\ll P_{XT}$ supplies overlap: combinations of
covariates and treatment with positive probability under the
target distribution must also be supported by the observed
distribution.

Our estimation task is to learn $w$ from the observed pairs
$\{(x_i,t_i)\}_{i=1}^N$ and use its estimate $\widehat w$ as
balancing weights. Outcomes enter the subsequent causal-response estimator.
We write $\widehat P_{XT}$, $\widehat P_X$, and $\widehat P_T$
for the empirical joint and marginal measures, and
$\widehat Q=\widehat P_X\otimes\widehat P_T$ for their product.
For measurable sets $A,B\subseteq\mathcal X\times\mathcal T$ with
$P_{XT}(A)>0$, define $P_A(B):=P_{XT}(B\cap A)/P_{XT}(A)$.
Similarly, $Q_A(B):=Q(B\cap A)/Q(A)$ when $Q(A)>0$.

\subsection{Cell-wise weight estimation}

A balancing tree estimates $w$ by comparing
joint and marginal frequencies within rectangular regions of
covariate--treatment space. Its leaves form a partition
$\pi=\{A_1,\ldots,A_L\}$ of $\mathcal X\times\mathcal T$, and
the estimated weight is constant within each leaf. Each cell has
the form $A=A_X\times A_T$, with continuous coordinates restricted
to intervals and categorical coordinates to subsets of their
categories.

The rectangular structure makes the reference probability factor:
$Q(A)=P_X(A_X)P_T(A_T)$. We therefore estimate the weight within
an occupied leaf by
\begin{equation}
    \label{eq:r-cell}
    \widehat w_A
    =
    \frac
         {\widehat P_X(A_X)\widehat P_T(A_T)}{\widehat P_{XT}(A)}
    =
    \frac{n_X(A_X)n_T(A_T)}{N\,n_{XT}(A)},
\end{equation}
where $n_{XT}(A)$ counts observed pairs in $A$, while
$n_X(A_X)$ and $n_T(A_T)$ count observations in its respective
projections. A weight below one indicates that the cell occurs more
frequently than expected under independence.

For arbitrary cell values $g_A\geq 0$, define
\begin{equation}
  \label{eq:partition-estimate}
  g_\pi(x, t) = \sum_{A \in \pi} \1\{(x,t) \in A\} g_A.
\end{equation}

The population cell-average approximation takes $g_A=w_A$:
\begin{equation}
  \label{eq:partition-population}
  w_\pi(x,t)=\sum_{A\in\pi}\1\{(x,t)\in A\}w_A,
  \qquad
  w_A:=\frac{Q(A)}{P_{XT}(A)}=\E_{P_{XT}}[w\mid (X,T)\in A],
\end{equation}
where cells with $P_{XT}(A)=0$ are omitted. Equivalently,
$w_\pi=\E_{P_{XT}}[w\mid\sigma(\pi)]$. The tree estimate takes
$g_A=\widehat w_A$ from \Cref{eq:r-cell}: for a pair $(x,t)$ in an
occupied leaf $A$, it returns $\widehat w_\pi(x,t)=\widehat w_A$.

\subsection{Learning the partition}

We grow the tree by successive binary refinements, starting from
$\pi^{(0)}=\{\mathcal X\times\mathcal T\}$. At iteration $k$,
a candidate split divides a leaf $A\in\pi^{(k)}$ into
two disjoint children $A_\ell$ and $A_r$. Each split acts on
one coordinate of either $X$ or $T$: continuous coordinates
are split at a threshold, while categorical coordinates are
split by membership in a subset of categories. Both children
therefore retain the rectangular structure of their parent.
The candidate partition is
\[
    \pi'
    =
    \bigl(\pi^{(k)}\setminus\{A\}\bigr)
    \cup\{A_\ell,A_r\}.
\]

For a differentiable, strictly convex function $f:(0,\infty)\to\mathbb R$ with
$f(1)=0$, the Bregman divergence between $a,b>0$ is
$B_f(a,b):=f(a)-f(b)-f'(b)(a-b)$, and the $f$-divergence between probability
measures $Q\ll P$ is $D_f(Q \mid\mid P):=\E_P[f(dQ/dP)]$.
We search for the partition and cell values that minimize the population risk
$R(g_\pi) := \E_{\PXT}[B_f(w,g_\pi)]$. \Cref{thm:Bregman-decomposition},
which follows the Bregman information decomposition of
\citet{banerjee2005clustering}, splits $R(g_\pi)$ into the divergence between
$Q$ and $P_{XT}$, which does not depend on the partition, and a term that does.

\begin{restatable}[Bregman decomposition \citep{banerjee2005clustering}]{theorem}{thmBregman}
  \label{thm:Bregman-decomposition}
  Let $P$ and $Q$ be probability measures on a common space with $Q\ll P$ and
  $D_f(Q \mid\mid P)<\infty$. Let $g_\pi$ be a cell-constant approximation of
  the Radon--Nikodym derivative $w = dQ/dP$ on a finite partition $\pi$, as
  defined in \Cref{eq:partition-estimate}, with $g_A>0$ and cells of positive
  mass under both measures.
  Let $P_\pi$ and $Q_\pi$ denote the measures induced on the cells of $\pi$, with
  $p_A := P(A)$, $q_A := Q(A)$ and $w_A := q_A/p_A$ for any cell $A \in \pi$.
  The population risk $R(g_\pi) :=  \E_P[B_f(w,g_\pi)]$
  satisfies
  \begin{equation}
    \label{eq:Bregman-decomposition}
    R(g_\pi) = D_f(Q \mid\mid P) - J(g_\pi), \quad 
    J(g_\pi) = D_f(Q_\pi \mid\mid P_\pi) - \sum_{A \in \pi}p_A B_f(w_A,g_A)
  \end{equation}
  with $D_f$ the $f$-divergence induced by $f$. In particular, if $g_A = w_A$ for all $A \in \pi$, 
  then:
  
  \begin{equation}
    \label{eq:Bregman-decomposition-optimal-g_A}
    R(w_\pi) = D_f(Q \mid\mid P) - D_f(Q_\pi \mid\mid P_\pi).  
  \end{equation}
  
\end{restatable}

The proof is given in \cref{app:proofs}. In the remainder, we apply the theorem
with $P=P_{XT}$ and $Q=P_X\otimes P_T$. \Cref{thm:Bregman-decomposition} shows
that the population risk is minimized by maximizing the $f$-divergence between
the induced measures $Q_\pi$ and $P_\pi$.
This motivates the following definition of the gain of a split. Let $g_{\pi'}$ agree with
$g_\pi$ outside $A$ and take the values $g_{A_\ell}$ and $g_{A_r}$ on the children.
Since every other cell keeps its contribution, \Cref{eq:Bregman-decomposition} gives
the gain of the split:
\begin{align}
  \label{eq:gain}
  G(\pi, \pi') &= R(g_\pi) - R(g_{\pi'}) \nonumber \\
  &= \sum_{B \in \{A_\ell, A_r\}}p_B \big( f(w_B) - f(w_A) \big)
  + p_A B_f(w_A, g_A) - \sum_{B \in \{A_\ell, A_r\}} p_B B_f(w_B, g_B).
\end{align}
When $g_C = w_C$ for every cell $C$, the Bregman terms vanish and
\begin{equation}
  \label{eq:gain-optimal}
  G(\pi, \pi') = \sum_{B \in \{A_\ell, A_r\}}p_B \big( f(w_B) - f(w_A) \big).
\end{equation}

By Jensen's inequality, the gain in \cref{eq:gain-optimal} is nonnegative,
since $w_A$ is the average of $w_{A_\ell}$ and $w_{A_r}$ with weights
$p_{A_\ell}/p_A$ and $p_{A_r}/p_A$.
The choice of $f$ determines the gain criterion and how error is penalized.
We focus on two choices. \Cref{cor:mse-gain} shows that we can choose $f$ to minimize
the squared error between the true ratio $w$ and its cell-constant approximation $g_\pi$,
with a gain criterion that depends only on the cell statistics $p_A$, $q_A$ and $g_A$.

\begin{corollary}[Squared-error gain]
  \label[corollary]{cor:mse-gain}
  Let $f(t) = (t - 1)^2$. Then $B_f(a, b) = (a - b)^2$, $D_f(Q \mid\mid P) = \chi^2(Q \mid\mid P)$ and
  $D_f(Q_\pi \mid\mid P_\pi) = \sum_{A \in \pi}p_A(w_A - 1)^2 = \chi^2(Q_\pi \mid\mid P_\pi)$.
  The population Bregman divergence is the expected squared error of the weights,
  \begin{equation}
    \label{eq:mse-risk}
    R(g_\pi) = \E_P[(w - g_\pi)^2]
    = \chi^2(Q \mid\mid P) - \chi^2(Q_\pi \mid\mid P_\pi) + \sum_{A \in \pi} p_A (w_A - g_A)^2,
  \end{equation}
  and the gain of a split is
  \begin{equation}
    \label{eq:mse-gain}
    G_{L^2}(\pi, \pi') = \sum_{B \in \{A_\ell, A_r\}}p_B (w_B - 1)^2 - p_A(w_A - 1)^2
    + p_A (w_A - g_A)^2 - \sum_{B \in \{A_\ell, A_r\}} p_B (w_B - g_B)^2.
  \end{equation}
  In particular, if $g_C = w_C$ for every cell $C$, the gain is the increase in the
  chi-squared divergence between the induced measures:
  \begin{equation}
    \label{eq:mse-gain-optimal}
    G_{L^2}(\pi, \pi') = \sum_{B \in \{A_\ell, A_r\}}p_B (w_B - 1)^2 - p_A(w_A - 1)^2.
  \end{equation}
\end{corollary}

When $\chi^2(Q\mid\mid P_{XT})<\infty$, define the squared-error
partition residual by
\begin{equation}
  \label{eq:residual-l2}
  \Delta_\pi^{L^2}
  :=\chi^2(Q\mid\mid P_{XT})-\chi^2(Q_\pi\mid\mid P_\pi)
  =\E_{\PXT}[(w-w_\pi)^2].
\end{equation}

Since $w_\pi$ has mean one under $P_{XT}$, the chi-squared divergence
$\chi^2(Q_\pi \mid\mid P_\pi)$ equals the variance of the weights $\mathrm{Var}_{\PXT}[w_\pi]$.

\Cref{cor:mi-gain} shows that, with $f = -\log$, the gain equals
the increase in the mutual information between $X$ and $T$ captured by the partition.

\begin{corollary}[Mutual-information gain]
  \label[corollary]{cor:mi-gain}
  Let $P = P_{XT}$, $Q = P_X \otimes P_T$ and $f(t) = -\log t$. Then
  $B_f(a, b) = a/b - \log(a/b) - 1$ is the Itakura--Saito divergence,
  $D_f(Q \mid\mid P) = D_{\mathrm{KL}}(P \mid\mid Q) = I(X;T)$, and
  $D_f(Q_\pi \mid\mid P_\pi) = \sum_{A \in \pi} p_A \log(p_A/q_A) = D_{\mathrm{KL}}(P_\pi \mid\mid Q_\pi) =: I_\pi$
  is the mutual information captured by the partition.
  The population Bregman divergence is
  \begin{equation}
    \label{eq:mi-risk}
    R(g_\pi) = \E_P\left[\frac{w}{g_\pi} - \log\frac{w}{g_\pi} - 1\right]
    = I(X;T) - I_\pi + \sum_{A \in \pi} p_A \left(\frac{w_A}{g_A} - \log\frac{w_A}{g_A} - 1\right),
  \end{equation}
  and the gain of a split is
  \begin{equation}
    \label{eq:mi-gain}
    G_{\mathrm{MI}}(\pi, \pi') = \sum_{B \in \{A_\ell, A_r\}}p_B \log\frac{p_B}{q_B} - p_A \log\frac{p_A}{q_A}
    + p_A B_f(w_A, g_A) - \sum_{B \in \{A_\ell, A_r\}} p_B B_f(w_B, g_B).
  \end{equation}
  In particular, if $g_C = w_C$ for every cell $C$, the gain is the increase in the
  mutual information captured by the partition:
  \begin{equation}
    \label{eq:mi-gain-optimal}
    G_{\mathrm{MI}}(\pi, \pi') = I_{\pi'} - I_\pi
    = \sum_{B \in \{A_\ell, A_r\}}p_B \log\frac{p_B}{q_B} - p_A \log\frac{p_A}{q_A}.
  \end{equation}
\end{corollary}

When $I(X;T)<\infty$, define the residual mutual information by
\begin{equation}
  \label{eq:residual-mi}
  \Delta_\pi^{\mathrm{MI}}:=I(X;T)-I_\pi
  =\E_{\PXT}\left[
    \frac{w}{w_\pi}-\log\frac{w}{w_\pi}-1
  \right].
\end{equation}
Both residuals $\Delta_\pi^{\mathrm{MI}}$ and $\Delta_\pi^{L^2}$ measure population partition approximation error,
while the cell-estimation terms in \cref{eq:mse-risk,eq:mi-risk} account for
the use of fitted cell values.

\subsection{Split search}

The gains of \cref{cor:mse-gain,cor:mi-gain} require the masses $p_B$ and
$q_B$ of each candidate child $B$. We use their empirical counterparts
$\widehat p_B$ and $\widehat q_B$, computed from the counts $n_{XT}$, $n_X$
and $n_T$ of \cref{eq:r-cell}, and evaluate the gains with $g_C=w_C$, that is,
\cref{eq:mse-gain-optimal,eq:mi-gain-optimal}. The search follows Partition
Tree \citep{angelim2026partition}.

\paragraph{Ordered coordinates.}
We sort each ordered coordinate once and visit its candidate thresholds in
increasing order, updating the three counts of the left child cumulatively.
Evaluating a coordinate therefore takes a single scan, as in decision trees
\citep{breiman1984classification}. To reduce the cost further, we restrict
thresholds to the boundaries of at most $H$ quantile bins per coordinate and
scan bin histograms, as in histogram-based gradient boosting
\citep{ke2017lightgbm}.

\paragraph{Categorical coordinates.}
For a categorical coordinate $j$, let $A_{j,c}$ denote the part of the leaf
whose coordinate $j$ equals category $c$. We sort the categories by the ratio
$\widehat p_{A_{j,c}}/\widehat q_{A_{j,c}}$ and evaluate each prefix of the
sorted categories as a candidate left child, following Partition Tree
\citep{angelim2026partition}.

\paragraph{Exploration splits.}
Every single-coordinate split of the root has zero gain. A child of the root
has the form $A_X\times\mathcal T$ or $\mathcal X\times A_T$, so its joint
mass equals its product mass and its weight is one. When no candidate has
positive gain, the tree takes an exploration split, which divides a leaf into
two children with balanced joint counts along one coordinate, alternating
between covariate and treatment coordinates. A tree takes at most $k_e$
exploration splits, and positive-gain splits always take priority.

\paragraph{Stopping.}
A child is admissible only if its joint count reaches a tuned minimum and
each marginal count is at least $\kappa\sqrt{N}$, where $\kappa>0$ is the tuned
marginal floor factor and $N$ is the training sample size. Growth stops when
no admissible split has gain above a minimum, or at a maximum
number of leaves. \Cref{app:split-search} gives the details of the search.

\subsection{Balancing Partition Forests}
A Balancing Partition Forest grows $N_F$ trees, each on a bootstrap resample
of the training pairs. The forest averages the joint-to-product ratios
$1/\widehat w_\pi$ of its trees, which is the harmonic mean of their weights.
For a set of trees $F = \{\pi_1,\ldots,\pi_{N_F}\}$, the forest estimate is
\begin{equation}
  \widehat w_{F}(x,t) = \left(\frac{1}{N_F}\sum_{\pi \in F} \widehat w_\pi(x,t)^{-1}\right)^{-1}.
\end{equation}
The harmonic mean is at most the geometric and arithmetic means, so a single
tree with a large cell weight cannot dominate the forest weight.

\section{Causal bias bounds}\label{sec:properties}

We relate the partition residuals to population causal-response bias.
Let $m(x,t)=\E[Y\mid X=x,T=t]$ denote the conditional outcome mean
and $\widehat m(x,t)$ an approximation. For the population analysis,
treat the nuisance approximations as fixed and define the outcome error by
\begin{equation}
  \label{eq:outcome-error}
  e(x,t):=m(x,t)-\widehat m(x,t).
\end{equation}
The doubly robust pseudo-outcome \citep{kennedy2017nonparametric} is
\begin{equation}
  \label{eq:dr-estimator}
  \xi(\widehat w,\widehat m)
  =\widehat w(X,T)(Y-\widehat m(X,T))
   +\int\widehat m(x,T)\,dP_X(x).
\end{equation}
Under consistency, conditional exchangeability and positivity,
$\E[\xi(\widehat w,\widehat m)\mid T=t]=\mu(t)$ whenever either
$\widehat m=m$ or $\widehat w=w$, provided the expectations exist.
For the population partition weights, the conditional bias is
\begin{equation}
  \label{eq:partition-bias}
  \begin{aligned}
    b_\pi(t)
    &:=\E[\xi(w_\pi,\widehat m)\mid T=t]-\mu(t)\\
    &=\E_{P_X}\left[
      \frac{w_\pi(X,t)-w(X,t)}{w(X,t)}e(X,t)
    \right]\\
    &=\E_{\PXT}[(w_\pi-w)e\mid T=t].
  \end{aligned}
\end{equation}
The second equality takes the conditional expectation of
$Y-\widehat m(X,T)$ given $(X,T)$, writes $\mu(t)=\int m(x,t)\,dP_X(x)$ by
consistency and conditional exchangeability, and uses
$dP_{X\mid T=t}=w(X,t)^{-1}dP_X$ for $P_T$-almost every $t$.
The third applies the same change of measure in reverse.
Pure IPW corresponds to $\widehat m\equiv0$, in which case $e=m$.
The bias $b_\pi$ concerns the population weights $w_\pi$ on a fixed partition.

\begin{restatable}[Causal bias bounds from the squared-error residual]{proposition}{thmbiasltwo}
  \label{prop:bias-bound-l2}
  Suppose $\chi^2(Q\mid\mid P_{XT})<\infty$ and
  $\E_{\PXT}[e^2]<\infty$. Then
  \begin{equation}
    \label{eq:bias-bound-l2}
    \E_{P_T}[|b_\pi(T)|]
    \leq\sqrt{\Delta_\pi^{L^2}}\sqrt{\E_{\PXT}[e^2]}.
  \end{equation}
  If additionally $|e|\leq M$ almost surely, then
  \begin{equation}
    \label{eq:squared-bias-bound-l2}
    \E_{P_T}[b_\pi(T)^2]\leq M^2\Delta_\pi^{L^2}.
  \end{equation}
\end{restatable}

\begin{lemma}[Total variation and absolute weight error]
  \label[lemma]{lem:mae-tv}
  For probability measures $U,V$ with $V\ll U$, let
  $\mathrm{TV}(U,V):=\frac12\int|1-dV/dU|\,dU$ denote the total variation
  distance. The expected absolute error of $w_\pi$ satisfies
  \begin{equation}
    \label{eq:mae-tv}
    \E_{\PXT}[| w_\pi - w |] = 2 \sum_{A \in \pi} q_A \mathrm{TV}(P_A, Q_A)
     \leq \sum_{A \in \pi} q_A \sqrt{2D_{\mathrm{KL}}(P_A \mid\mid Q_A)}.
  \end{equation}
\end{lemma}

The next result connects causal bias to the residual mutual information.
The proofs of this section are given in \cref{app:proofs}.

\begin{restatable}[Causal bias bounds from residual mutual information]{proposition}{thmbiasmi}
  \label{prop:bias-bound-mi}
  Suppose $I(X;T)<\infty$ and $|e|\leq M$ almost surely. Then
  \begin{equation}
    \label{eq:bias-bound-mi}
    \E_{P_T}[|b_\pi(T)|]
    \leq M\sqrt{2\E_{\PXT}[w_\pi^2]\Delta_\pi^{\mathrm{MI}}}.
  \end{equation}
  If additionally $0<w\leq W$ almost surely, then
  \begin{align}
    \E_{P_T}[|b_\pi(T)|]
    &\leq M\sqrt{2W\Delta_\pi^{\mathrm{MI}}},
      \label{eq:bias-bound-mi-overlap}\\
    \E_{P_T}[b_\pi(T)^2]
    &\leq2M^2W^2\Delta_\pi^{\mathrm{MI}}.
      \label{eq:squared-bias-bound-mi}
  \end{align}
\end{restatable}

\section{Experiments}\label{sec:experiments}

\subsection{Setup and estimators}

We use a common experimental protocol across all weight estimators, holding
the datasets, seeds, hyperparameter budget and causal-response estimators fixed.
Therefore, comparisons within a causal-response estimator differ only in the
estimated weights.

We evaluate seven synthetic data-generating processes: the Kang--Schafer
design \citep{10.1214/07-STS227} with binary treatment, a
continuous-treatment version of it, a misspecified variant of each, a
Beta-mixture design that we introduce, with binary and continuous
treatment, and SyntheticVCNet \citep{nie2021vcnet}.
We use training sizes $N\in\{500,2{,}000\}$ and five seeds. For each seed, we
draw an independent evaluation sample of $10{,}000$ observations from the same
data-generating process used for training.

We also consider three semi-synthetic settings based on the covariates of
three datasets: Digits ($1{,}797$ observations, 64 covariates), ACIC~2016 ($4{,}802$, 58)
and California housing ($20{,}640$, 8). We generate a continuous treatment and an
outcome using nonlinear random functions of the observed covariates. We use
five outer folds, training on four folds and evaluating on the remaining fold.
The seed determines both the simulated data-generating process and the held-out
fold. Hyperparameter selection uses only the corresponding training sample. We
give the complete data-generating processes in \cref{app:dgp} and the tuning
protocol and search spaces in \cref{app:search-spaces}.
We consider the following weight estimators:
\begin{itemize}
  \item \textbf{Balancing Partition Tree and Balancing Partition Forest (BPT and BPF)}. BPT and BPF use the
  mutual-information gain in \cref{eq:mi-gain-optimal}, while
  BPT$_{L^2}$ and BPF$_{L^2}$ use the squared-error gain in
  \cref{eq:mse-gain-optimal}. All variants return the cell weight of
  \cref{eq:r-cell}, and pairs outside the
  training support receive a weight of one. BPF and BPF$_{L^2}$ combine
  50 bootstrap trees. We tune the marginal and joint count floors.
  \item \textbf{Permutation Weighting with Gradient Boosting (PW-GB)}
  \citep{arbour2021permutation}. A histogram gradient-boosting classifier
  distinguishes observed pairs from pairs with permuted treatment; we average
  probabilities over 100 permutations to estimate the weight.
  \item \textbf{Conditional probability model (Cond.)}. Histogram gradient
  boosting estimates treatment probabilities for binary treatment and a
  Gaussian conditional density for continuous treatment.
  \item \textbf{CBPS} \citep{imai2014covariate,fong2018covariate}. We use
  the overidentified covariate-balancing propensity score estimator with
  multinomial-logit and Gaussian treatment models for categorical and
  continuous treatments, respectively.
\end{itemize}
We only fit estimators that support the treatment type of the dataset.
Implementation details appear in \cref{app:weight-estimators}. The experimental workflow follows two steps:
\begin{enumerate}
  \item \textbf{Weight estimation}. We select each tunable estimator by
  three-fold cross-validation using the held-out permutation log loss of
  \citet{arbour2021permutation}. All four partition variants use this same selection criterion,
  irrespective of their split objective. We use 25 trials per tunable model;
  \cref{app:search-spaces} gives the loss and search spaces.
  \item \textbf{Causal-response estimation}. We combine the weight and
  outcome models using three response estimators:
  \begin{enumerate}
    \item A Marginal Structural Model (MSM) \citep{robins2000marginal}, which
    fits a treatment-only cubic-spline ridge regression using three-fold
    out-of-fold weights.
    \item A Generalized Propensity Score (GPS) estimator \citep{hirano2004propensity},
    which regresses $Y$ on $(T,\widehat w(X,T))$ with an untuned histogram
    gradient-boosting model using three-fold out-of-fold scores. We average
    predictions over at most 500 covariate samples.
    \item Pseudo-outcome doubly robust estimation
    \citep{kennedy2017nonparametric}, as in \Cref{eq:dr-estimator}, using three outer folds. Within each
    fold, we fit the weight and outcome models on disjoint halves of the
    training sample and smooth the resulting pseudo-outcome over treatment.
    The outcome model $\widehat m(x,t)$ is a histogram gradient-boosting
    regressor whose configuration is selected by cross-validated $R^2$.
  \end{enumerate}
\end{enumerate}
We evaluate response RMSE at both treatment levels for binary treatment and on
a 100-point grid for continuous treatment; \cref{app:experimental_protocol}
specifies the grid and reference curve.
The detailed
response tables show the mean and sample standard deviation over five seeds or folds.
Ranks use the unrounded means.
Overall mean ranks in the response tables weight each dataset--sample-size
combination equally.

\subsection{Results}

\begin{table}[p]
\centering
\footnotesize
\setlength{\tabcolsep}{4pt}
\begin{adjustbox}{max width=\linewidth}
\begin{tabular}{lrrrrrrr}
\toprule
Dataset & BPT & BPF & BPT$_{L^2}$ & BPF$_{L^2}$ & PW-GB & CBPS & Cond. \\
\midrule
\multicolumn{8}{l}{\textit{Synthetic, $N=500$}} \\
KS binary & \makecell[r]{${2.803}$\\[-1.5pt]\color{black!65}$\pm 0.907$} & \makecell[r]{${\mathbf{1.809}}$\\[-1.5pt]\color{black!65}$\pm 0.947$} & \makecell[r]{${3.186}$\\[-1.5pt]\color{black!65}$\pm 0.953$} & \makecell[r]{${2.190}$\\[-1.5pt]\color{black!65}$\pm 1.256$} & \makecell[r]{${3.296}$\\[-1.5pt]\color{black!65}$\pm 1.895$} & \makecell[r]{${2.357}$\\[-1.5pt]\color{black!65}$\pm 1.176$} & \makecell[r]{${17.112}$\\[-1.5pt]\color{black!65}$\pm 6.014$} \\
KS binary, misspec. & \makecell[r]{${2.646}$\\[-1.5pt]\color{black!65}$\pm 0.929$} & \makecell[r]{${\mathbf{2.235}}$\\[-1.5pt]\color{black!65}$\pm 1.683$} & \makecell[r]{${2.269}$\\[-1.5pt]\color{black!65}$\pm 1.786$} & \makecell[r]{${2.632}$\\[-1.5pt]\color{black!65}$\pm 1.551$} & \makecell[r]{${4.205}$\\[-1.5pt]\color{black!65}$\pm 1.546$} & \makecell[r]{${7.513}$\\[-1.5pt]\color{black!65}$\pm 6.094$} & \makecell[r]{${25.864}$\\[-1.5pt]\color{black!65}$\pm 12.536$} \\
Beta-mixture binary & \makecell[r]{${0.220}$\\[-1.5pt]\color{black!65}$\pm 0.083$} & \makecell[r]{${0.140}$\\[-1.5pt]\color{black!65}$\pm 0.060$} & \makecell[r]{${0.230}$\\[-1.5pt]\color{black!65}$\pm 0.061$} & \makecell[r]{${0.162}$\\[-1.5pt]\color{black!65}$\pm 0.067$} & \makecell[r]{${0.225}$\\[-1.5pt]\color{black!65}$\pm 0.065$} & \makecell[r]{${\mathbf{0.076}}$\\[-1.5pt]\color{black!65}$\pm 0.033$} & \makecell[r]{${0.729}$\\[-1.5pt]\color{black!65}$\pm 0.436$} \\
KS continuous & \makecell[r]{${12.871}$\\[-1.5pt]\color{black!65}$\pm 2.089$} & \makecell[r]{${12.180}$\\[-1.5pt]\color{black!65}$\pm 1.013$} & \makecell[r]{${12.579}$\\[-1.5pt]\color{black!65}$\pm 1.765$} & \makecell[r]{${10.873}$\\[-1.5pt]\color{black!65}$\pm 1.589$} & \makecell[r]{${12.113}$\\[-1.5pt]\color{black!65}$\pm 1.866$} & \makecell[r]{${\mathbf{10.064}}$\\[-1.5pt]\color{black!65}$\pm 4.002$} & \makecell[r]{${13.614}$\\[-1.5pt]\color{black!65}$\pm 1.840$} \\
KS continuous, misspec. & \makecell[r]{${13.928}$\\[-1.5pt]\color{black!65}$\pm 1.901$} & \makecell[r]{${12.015}$\\[-1.5pt]\color{black!65}$\pm 0.896$} & \makecell[r]{${12.072}$\\[-1.5pt]\color{black!65}$\pm 2.722$} & \makecell[r]{${\mathbf{10.716}}$\\[-1.5pt]\color{black!65}$\pm 1.110$} & \makecell[r]{${14.397}$\\[-1.5pt]\color{black!65}$\pm 1.175$} & \makecell[r]{${30.229}$\\[-1.5pt]\color{black!65}$\pm 16.424$} & \makecell[r]{${17.468}$\\[-1.5pt]\color{black!65}$\pm 4.950$} \\
Beta-mixture continuous & \makecell[r]{${0.492}$\\[-1.5pt]\color{black!65}$\pm 0.215$} & \makecell[r]{${0.322}$\\[-1.5pt]\color{black!65}$\pm 0.166$} & \makecell[r]{${0.474}$\\[-1.5pt]\color{black!65}$\pm 0.219$} & \makecell[r]{${0.371}$\\[-1.5pt]\color{black!65}$\pm 0.177$} & \makecell[r]{${0.389}$\\[-1.5pt]\color{black!65}$\pm 0.131$} & \makecell[r]{${0.378}$\\[-1.5pt]\color{black!65}$\pm 0.086$} & \makecell[r]{${\mathbf{0.272}}$\\[-1.5pt]\color{black!65}$\pm 0.051$} \\
SyntheticVCNet & \makecell[r]{${0.136}$\\[-1.5pt]\color{black!65}$\pm 0.042$} & \makecell[r]{${\mathbf{0.113}}$\\[-1.5pt]\color{black!65}$\pm 0.029$} & \makecell[r]{${0.126}$\\[-1.5pt]\color{black!65}$\pm 0.012$} & \makecell[r]{${0.121}$\\[-1.5pt]\color{black!65}$\pm 0.029$} & \makecell[r]{${0.118}$\\[-1.5pt]\color{black!65}$\pm 0.029$} & \makecell[r]{${0.144}$\\[-1.5pt]\color{black!65}$\pm 0.059$} & \makecell[r]{${0.179}$\\[-1.5pt]\color{black!65}$\pm 0.078$} \\
\addlinespace[3pt]
Mean rank & 4.86 & 1.86 & 4.43 & 2.43 & 4.43 & 4.00 & 6.00 \\
\midrule
\multicolumn{8}{l}{\textit{Synthetic, $N=2{,}000$}} \\
KS binary & \makecell[r]{${1.897}$\\[-1.5pt]\color{black!65}$\pm 0.855$} & \makecell[r]{${1.441}$\\[-1.5pt]\color{black!65}$\pm 0.481$} & \makecell[r]{${1.785}$\\[-1.5pt]\color{black!65}$\pm 0.557$} & \makecell[r]{${1.456}$\\[-1.5pt]\color{black!65}$\pm 0.468$} & \makecell[r]{${2.077}$\\[-1.5pt]\color{black!65}$\pm 0.449$} & \makecell[r]{${\mathbf{1.061}}$\\[-1.5pt]\color{black!65}$\pm 0.662$} & \makecell[r]{${30.319}$\\[-1.5pt]\color{black!65}$\pm 18.128$} \\
KS binary, misspec. & \makecell[r]{${2.287}$\\[-1.5pt]\color{black!65}$\pm 0.911$} & \makecell[r]{${2.090}$\\[-1.5pt]\color{black!65}$\pm 0.685$} & \makecell[r]{${\mathbf{1.723}}$\\[-1.5pt]\color{black!65}$\pm 0.758$} & \makecell[r]{${2.095}$\\[-1.5pt]\color{black!65}$\pm 0.545$} & \makecell[r]{${3.497}$\\[-1.5pt]\color{black!65}$\pm 0.566$} & \makecell[r]{${3.997}$\\[-1.5pt]\color{black!65}$\pm 0.707$} & \makecell[r]{${26.521}$\\[-1.5pt]\color{black!65}$\pm 14.857$} \\
Beta-mixture binary & \makecell[r]{${0.106}$\\[-1.5pt]\color{black!65}$\pm 0.069$} & \makecell[r]{${0.081}$\\[-1.5pt]\color{black!65}$\pm 0.032$} & \makecell[r]{${0.134}$\\[-1.5pt]\color{black!65}$\pm 0.074$} & \makecell[r]{${0.093}$\\[-1.5pt]\color{black!65}$\pm 0.048$} & \makecell[r]{${0.141}$\\[-1.5pt]\color{black!65}$\pm 0.065$} & \makecell[r]{${\mathbf{0.052}}$\\[-1.5pt]\color{black!65}$\pm 0.028$} & \makecell[r]{${1.043}$\\[-1.5pt]\color{black!65}$\pm 0.942$} \\
KS continuous & \makecell[r]{${10.884}$\\[-1.5pt]\color{black!65}$\pm 3.300$} & \makecell[r]{${9.665}$\\[-1.5pt]\color{black!65}$\pm 0.834$} & \makecell[r]{${7.186}$\\[-1.5pt]\color{black!65}$\pm 1.796$} & \makecell[r]{${\mathbf{6.465}}$\\[-1.5pt]\color{black!65}$\pm 1.020$} & \makecell[r]{${8.291}$\\[-1.5pt]\color{black!65}$\pm 1.727$} & \makecell[r]{${8.511}$\\[-1.5pt]\color{black!65}$\pm 5.715$} & \makecell[r]{${13.376}$\\[-1.5pt]\color{black!65}$\pm 11.102$} \\
KS continuous, misspec. & \makecell[r]{${10.955}$\\[-1.5pt]\color{black!65}$\pm 3.256$} & \makecell[r]{${8.989}$\\[-1.5pt]\color{black!65}$\pm 0.649$} & \makecell[r]{${8.730}$\\[-1.5pt]\color{black!65}$\pm 2.504$} & \makecell[r]{${\mathbf{6.050}}$\\[-1.5pt]\color{black!65}$\pm 1.229$} & \makecell[r]{${10.252}$\\[-1.5pt]\color{black!65}$\pm 0.733$} & \makecell[r]{${22.977}$\\[-1.5pt]\color{black!65}$\pm 7.362$} & \makecell[r]{${9.274}$\\[-1.5pt]\color{black!65}$\pm 1.882$} \\
Beta-mixture continuous & \makecell[r]{${0.283}$\\[-1.5pt]\color{black!65}$\pm 0.139$} & \makecell[r]{${\mathbf{0.181}}$\\[-1.5pt]\color{black!65}$\pm 0.108$} & \makecell[r]{${0.297}$\\[-1.5pt]\color{black!65}$\pm 0.159$} & \makecell[r]{${0.201}$\\[-1.5pt]\color{black!65}$\pm 0.117$} & \makecell[r]{${0.235}$\\[-1.5pt]\color{black!65}$\pm 0.064$} & \makecell[r]{${0.470}$\\[-1.5pt]\color{black!65}$\pm 0.238$} & \makecell[r]{${0.382}$\\[-1.5pt]\color{black!65}$\pm 0.211$} \\
SyntheticVCNet & \makecell[r]{${\mathbf{0.073}}$\\[-1.5pt]\color{black!65}$\pm 0.037$} & \makecell[r]{${0.083}$\\[-1.5pt]\color{black!65}$\pm 0.016$} & \makecell[r]{${0.075}$\\[-1.5pt]\color{black!65}$\pm 0.012$} & \makecell[r]{${0.075}$\\[-1.5pt]\color{black!65}$\pm 0.021$} & \makecell[r]{${0.098}$\\[-1.5pt]\color{black!65}$\pm 0.021$} & \makecell[r]{${0.161}$\\[-1.5pt]\color{black!65}$\pm 0.041$} & \makecell[r]{${0.189}$\\[-1.5pt]\color{black!65}$\pm 0.073$} \\
\addlinespace[3pt]
Mean rank & 4.29 & 2.71 & 3.14 & 2.14 & 4.71 & 4.57 & 6.43 \\
\midrule
\multicolumn{8}{l}{\textit{Semi-synthetic}} \\
Digits & \makecell[r]{${0.778}$\\[-1.5pt]\color{black!65}$\pm 0.306$} & \makecell[r]{${\mathbf{0.672}}$\\[-1.5pt]\color{black!65}$\pm 0.246$} & \makecell[r]{${0.781}$\\[-1.5pt]\color{black!65}$\pm 0.312$} & \makecell[r]{${0.690}$\\[-1.5pt]\color{black!65}$\pm 0.262$} & \makecell[r]{${0.739}$\\[-1.5pt]\color{black!65}$\pm 0.291$} & \makecell[r]{${1.167}$\\[-1.5pt]\color{black!65}$\pm 0.671$} & \makecell[r]{${0.676}$\\[-1.5pt]\color{black!65}$\pm 0.242$} \\
ACIC & \makecell[r]{${0.782}$\\[-1.5pt]\color{black!65}$\pm 0.125$} & \makecell[r]{${0.527}$\\[-1.5pt]\color{black!65}$\pm 0.162$} & \makecell[r]{${0.796}$\\[-1.5pt]\color{black!65}$\pm 0.112$} & \makecell[r]{${0.523}$\\[-1.5pt]\color{black!65}$\pm 0.160$} & \makecell[r]{${0.502}$\\[-1.5pt]\color{black!65}$\pm 0.142$} & \makecell[r]{${\mathbf{0.458}}$\\[-1.5pt]\color{black!65}$\pm 0.367$} & \makecell[r]{${4.685}$\\[-1.5pt]\color{black!65}$\pm 6.110$} \\
California & \makecell[r]{${0.501}$\\[-1.5pt]\color{black!65}$\pm 0.234$} & \makecell[r]{${0.410}$\\[-1.5pt]\color{black!65}$\pm 0.168$} & \makecell[r]{${0.553}$\\[-1.5pt]\color{black!65}$\pm 0.182$} & \makecell[r]{${0.447}$\\[-1.5pt]\color{black!65}$\pm 0.161$} & \makecell[r]{${0.420}$\\[-1.5pt]\color{black!65}$\pm 0.186$} & \makecell[r]{${6.785}$\\[-1.5pt]\color{black!65}$\pm 11.895$} & \makecell[r]{${\mathbf{0.319}}$\\[-1.5pt]\color{black!65}$\pm 0.166$} \\
\addlinespace[3pt]
Mean rank & 5.00 & 2.33 & 6.00 & 3.33 & 3.00 & 5.00 & 3.33 \\
\midrule
Overall mean rank & 4.65 & 2.29 & 4.18 & 2.47 & 4.29 & 4.41 & 5.71 \\
\bottomrule
\end{tabular}
\end{adjustbox}
\caption{MSM response RMSE (mean $\pm$ sample SD, five seeds/folds). Bold marks the smallest mean. Mean ranks use unrounded means, midranks for ties, and equal weights for settings.}
\label{tab:causal_msm_response}

\end{table}

\begin{table}[p]
\centering
\footnotesize
\setlength{\tabcolsep}{4pt}
\begin{adjustbox}{max width=\linewidth}
\begin{tabular}{lrrrrrrr}
\toprule
Response estimator & BPT & BPF & BPT$_{L^2}$ & BPF$_{L^2}$ & PW-GB & CBPS & Cond. \\
\midrule
MSM & 4.65 & 2.29 & 4.18 & 2.47 & 4.29 & 4.41 & 5.71 \\
GPS & 5.65 & 2.12 & 5.71 & 2.47 & 2.59 & 4.06 & 5.41 \\
DR & 3.29 & 2.24 & 3.94 & 2.71 & 3.47 & 5.65 & 6.71 \\
\bottomrule
\end{tabular}
\end{adjustbox}
\caption{Overall mean response-RMSE ranks across 17 equally weighted settings; lower is better.}
\label{tab:causal_response_summary}

\end{table}

\paragraph{Causal-response accuracy.}
BPF achieves the lowest overall mean MSM RMSE rank across the
17 dataset--sample-size settings: 2.29, compared with 4.29 for PW-GB.
\Cref{tab:causal_msm_response} reports MSM RMSE means and
standard deviations for all 17 settings.

BPF has lower mean MSM RMSE than PW-GB in 14 of 17 settings.
For example, on misspecified binary Kang--Schafer at $N=500$,
BPF obtains $2.235 \pm 1.683$, compared with
$4.205 \pm 1.546$ for PW-GB.
Many pairwise differences are within one standard deviation across the five seeds.
CBPS remains competitive in several binary settings, achieving
the lowest MSM error on Beta-mixture binary at both sample sizes.

BPF also achieves the lowest overall mean RMSE rank under GPS and DR,
although the ordering varies by dataset and response estimator.
\Cref{app:response-results} reports the full GPS and DR results and
analyzes their accuracy, forest gains and weight-related instability.

\paragraph{Forests and split objectives.}
Forest aggregation generally improves MSM response accuracy over
individual partition trees: BPF improves on BPT
in 16 of 17 settings, and BPF$_{L^2}$ improves on
BPT$_{L^2}$ in 15 of 17.
Although BPF has a better overall mean MSM rank than BPF$_{L^2}$,
the squared-error objective offers substantial gains in some settings.
On continuous Kang--Schafer at $N=2{,}000$,
BPF$_{L^2}$ reduces mean MSM RMSE from 9.665 to 6.465;
on its misspecified counterpart, it reduces RMSE from
8.989 to 6.050.

\paragraph{Forest aggregation.}
We compare harmonic, geometric and arithmetic means of the tree weights,
independently retuning each variant.
Harmonic averaging has the lowest mean rank for MSM response RMSE under
both split objectives (\cref{tab:forest_ablation}).
\Cref{app:forest-ablation} gives the protocol and results.

\paragraph{Computational cost.}
We time fits on nested California subsamples ($d=8$) of 500 to 16,512 rows,
fixing each method's selected configuration from the first training fold.
At $N=16{,}512$, single-thread median fit times are 0.142 seconds for BPT,
13.004 for BPF, 0.136 for CBPS and 107.478 for PW-GB. With four threads, they
are 0.091, 3.259, 0.137 and 85.293 seconds.
PW-GB fits one classifier per permutation, and its time covers the 100
permutations of the selected configuration.
The timing protocol, scaling curves (\cref{fig:fit_time_vs_n}) and aggregate
refit timings (\cref{tab:causal_refit_times}) appear in \cref{app:computation}.

\section{Conclusion}

Estimating balancing weights requires capturing treatment--covariate dependence
while keeping estimation computationally practical.
We introduced Balancing Partition Tree and Balancing Partition Forest,
which estimate the balancing weight $d(P_X\otimes P_T)/dP_{XT}$ on learned
rectangular partitions. Each split maximizes a Bregman divergence gain
computed from joint and marginal counts.
The population residuals bound the absolute and squared bias of
inverse-probability-weighted and doubly robust estimators that use the
population partition weights.
In the experiments, BPF obtained the lowest mean RMSE rank under all three
response estimators. These findings support adaptive partition forests as a
practical approach to balancing-weight estimation for binary and continuous
treatments.
Future work includes statistical guarantees for learned partitions and forest
aggregation, and applications of BPT to interpretable balancing-weight
estimation in real-world settings.

\clearpage
\bibliography{references}

\appendix
\clearpage
\crefalias{section}{appendix}
\crefalias{subsection}{appendix}

\renewcommand{\contentsname}{Appendix contents}
\addtocontents{toc}{\protect\setcounter{tocdepth}{2}}
\tableofcontents

\section{Proofs}\label{app:proofs}

\thmBregman*

\begin{proof}
  Expand the Bregman divergence and use $\E_P[f(w)]=D_f(Q \mid\mid P)$:

\begin{align*}
  \mathbb{E}_P[B_f(w, g_\pi)] &= \E_P[f(w)] - \E_P[f(g_\pi)] - \E_P[f'(g_\pi)(w - g_\pi)] \\
  &=  D_f(Q \mid\mid P) - (\E_P[f(g_\pi) - f'(g_\pi)(g_\pi - w)]) \\
  &=  D_f(Q \mid\mid P) - (\E_P[f(g_\pi) - f'(g_\pi)g_\pi] + \E_Q[f'(g_\pi)]) \\
  &=  D_f(Q \mid\mid P) - \sum_{A \in \pi}\Big(p_A\left(f(g_A) - f'(g_A)g_A\right) + q_Af'(g_A)\Big) \\
  &=  D_f(Q \mid\mid P) - \sum_{A \in \pi}\Big(p_A\left(f(g_A) - f'(g_A)g_A\right) + p_Aw_Af'(g_A)\Big) \\
  &=  D_f(Q \mid\mid P) - \sum_{A \in \pi}\Big(p_A\left(f(g_A) - f'(g_A)g_A\right) + p_Aw_Af'(g_A) \\
  &\qquad + p_Af(w_A) - p_Af(w_A)\Big) \\
  &= D_f(Q \mid\mid P) - D_f(Q_\pi \mid\mid P_\pi) + \sum_{A \in \pi}p_A B_f(w_A,g_A)
\end{align*}

The last equality follows from the definitions of the $f$-divergence and the Bregman divergence.
Since $B_f(w_A,w_A) = 0$, the last term vanishes when $g_A = w_A$ for all $A \in \pi$.

\end{proof}

\thmbiasltwo*

\begin{proof}
Conditional Jensen's inequality followed by the Cauchy--Schwarz inequality gives
\begin{align*}
  \E_{P_T}[|b_\pi(T)|]
  &\leq\E_{\PXT}[|w_\pi-w|\,|e|]\\
  &\leq\sqrt{\E_{\PXT}[(w_\pi-w)^2]}
        \sqrt{\E_{\PXT}[e^2]}
   =\sqrt{\Delta_\pi^{L^2}}\sqrt{\E_{\PXT}[e^2]}.
\end{align*}
For squared bias, apply conditional Jensen's inequality before integrating:
\[
  b_\pi(t)^2
  \leq\E_{\PXT}[(w_\pi-w)^2e^2\mid T=t]
  \leq M^2\E_{\PXT}[(w_\pi-w)^2\mid T=t].
\]
Integration over $P_T$ and \cref{eq:residual-l2} give the second bound.
\end{proof}

\begin{proof}\textbf{(of \cref{lem:mae-tv})}
  For each cell $A\in\pi$,
  \begin{align*}
    \int_A | w_\pi - w | \,d{\PXT} &= p_A \int | w_A - w | \,dP_A
    = p_A w_A \int \Bigl| 1 - \frac{w}{w_A} \Bigr| \,dP_A \\
    &= q_A \int \Bigl| 1 - \frac{dQ_A}{dP_A}\Bigr| \,dP_A
    = 2q_A \mathrm{TV}(P_A, Q_A)
    \leq q_A\sqrt{2D_{\mathrm{KL}}(P_A \mid\mid Q_A)},
  \end{align*}
  where the inequality is Pinsker's inequality.
  Summing over $A \in \pi$ gives the result.
\end{proof}

\thmbiasmi*

\begin{proof}
The KL chain rule on the partition gives
\begin{align*}
  \sum_{A\in\pi}p_A D_{\mathrm{KL}}(P_A\mid\mid Q_A)
  &=\sum_{A\in\pi}\int_A
    \left[\log\frac{dP_{XT}}{dQ}-\log\frac{p_A}{q_A}\right]dP_{XT}\\
  &=I(X;T)-I_\pi=\Delta_\pi^{\mathrm{MI}}.
\end{align*}
By \cref{eq:mae-tv} and the Cauchy--Schwarz inequality,
\begin{align*}
  \E_{\PXT}[|w_\pi-w|]
  &\leq\sqrt{2}\sum_{A\in\pi}
    \frac{q_A}{\sqrt{p_A}}
    \sqrt{p_A D_{\mathrm{KL}}(P_A\mid\mid Q_A)}\\
  &\leq\sqrt{2\left(\sum_{A\in\pi}\frac{q_A^2}{p_A}\right)
                   \Delta_\pi^{\mathrm{MI}}}
   =\sqrt{2\E_{\PXT}[w_\pi^2]\Delta_\pi^{\mathrm{MI}}}.
\end{align*}
Since $\E_{P_T}[|b_\pi(T)|]\leq M\E_{\PXT}[|w_\pi-w|]$,
this proves \cref{eq:bias-bound-mi}.
If $w\leq W$, then $w_\pi=\E_{\PXT}[w\mid\sigma(\pi)]\leq W$ and
$\E_{\PXT}[w_\pi^2]\leq W\E_{\PXT}[w_\pi]=W$, since $\E_{\PXT}[w_\pi] = 1$, giving
\cref{eq:bias-bound-mi-overlap}.

For squared bias, $f(u)=-\log u$ satisfies $f''(u)=u^{-2}\geq W^{-2}$
on $(0,W]$. For $a,b\in(0,W]$, Taylor's formula gives
\[
  f(a) = f(b) + f'(b)(a-b) + \frac{1}{2}f''(\eta)(a-b)^2
\]
for some $\eta$ between $a$ and $b$, so that
\[
B_f(a, b) = f(a) - f(b) - f'(b)(a-b) \geq \frac{1}{2W^2}(a-b)^2.
\]
Setting $a=w$ and $b=w_\pi$, taking expectations and using
\cref{eq:residual-l2,eq:residual-mi} gives
\[
\Delta_\pi^{L^2}\leq2W^2\Delta_\pi^{\mathrm{MI}}.
\]
The squared-error residual is finite because $w\leq W$.
Applying \cref{eq:squared-bias-bound-l2} proves
\cref{eq:squared-bias-bound-mi}.
\end{proof}

\section{Implementation of the split search}\label{app:split-search}

BPT and each member of BPF use the same binary partition builder. This
section describes how it constructs and evaluates candidates for ordered
and categorical coordinates. The two gain objectives use the same candidate
sets and child statistics.

\subsection{Joint and marginal statistics}

For a cell $A=A_X\times A_T$, the builder maintains three collections of
training-row indices:
\begin{equation}
  \mathcal I_{XT}(A)=\{i:(X_i,T_i)\in A\},\qquad
  \mathcal I_X(A)=\{i:X_i\in A_X\},\qquad
  \mathcal I_T(A)=\{i:T_i\in A_T\}.
\end{equation}
The marginal collections are not restricted to the joint collection. For
example, $\mathcal I_X(A)$ includes rows whose treatments lie outside $A_T$.
During forest growth, these collections retain bootstrap multiplicities.
Write $n_{XT}(B)$, $n_X(B)$ and $n_T(B)$ for their counts in a cell $B$, and
$N_0$ for the number of row occurrences at the root of the growing tree, so
that $N_0=N$ for a single tree. Candidate scores use
\begin{equation}
  \widehat p_B=\frac{n_{XT}(B)}{N_0},\qquad
  \widehat q_B=\frac{n_X(B)n_T(B)}{N_0^2}.
  \label{eq:split-empirical-masses}
\end{equation}
An $X$-split divides the joint and $X$ collections, while both children
inherit the parent's $T$ collection. A $T$-split divides the joint and $T$
collections and preserves the $X$ collection. Thus, reference masses are
computed from two marginal counts without constructing the Cartesian
product of the samples. These statistics choose the partition. The fitted
cell weights use the counts of the full training sample, as in
\cref{eq:r-cell}.

The score of a candidate $A\to\{A_\ell,A_r\}$ is
\begin{align}
  \widehat G_{\mathrm{MI}}
  &=\sum_{B\in\{A_\ell,A_r\}}
    \widehat p_B\log\frac{\widehat p_B}{\widehat q_B}
    -\widehat p_A\log\frac{\widehat p_A}{\widehat q_A},\\
  \widehat G_{L^2}
  &=\sum_{B\in\{A_\ell,A_r\}}
    \frac{\widehat q_B^2}{\widehat p_B}
    -\frac{\widehat q_A^2}{\widehat p_A}.
  \label{eq:split-implemented-gains}
\end{align}
The second expression equals the empirical gain in
\cref{eq:mse-gain-optimal}: expanding $p(w-1)^2$ with $w=q/p$ leaves
$q^2/p$, since the terms linear in $p$ and $q$ cancel between parent and
children. Both children must contain joint observations and satisfy the
joint and marginal count floors. Candidates with non-finite gain
are discarded.

\subsection{Histogram boundaries for ordered coordinates}

A bin limit $H$ controls the resolution of each ordered coordinate. Each
ordered coordinate is sorted once on the training data, and consecutive equal
values are grouped into indivisible runs. If there are at most $H$ distinct
values, each run receives its own bin. Otherwise, bins combine consecutive
runs using approximately equal-frequency boundaries: for bin
$b=1,\ldots,H$, the desired cumulative row count is $\lceil bN/H\rceil$.
Starting from the next unassigned run, the bin extends through the first run
that reaches this rank. If the number of remaining runs is at most the number
of remaining bins, each remaining run receives its own bin. Ties therefore
stay together, and the resulting bins need not have equal sizes or widths.

Each row stores a bin identifier, and each bin stores its smallest and
largest observed values. The bins are fitted on the training sample supplied
to a fit, before bootstrap sampling, and remain fixed throughout growth. BPF
shares these bin assignments across its members.

At a node, an $X$ coordinate uses the bins occupied by
$\mathcal I_X(A)$, and a $T$ coordinate uses those occupied by
$\mathcal I_T(A)$. For each pair of consecutive occupied bins, the search
proposes one cut separating the lower bins from the upper bins. Empty bins
between them are skipped. For continuous values, the threshold is the
midpoint between the upper endpoint of the lower bin and the lower endpoint
of the upper bin. Integer coordinates use the lower endpoint of the upper bin
with a strict less-than comparison. Thus a
coordinate with $K$ occupied bins supplies at most $K-1\leq H-1$ cuts.

The experiments use $H=1{,}024$ for every ordered coordinate and no additional
cap on the number of candidate cuts. When every distinct training value
receives its own bin, histogram search retains the exact candidate boundaries
between consecutive distinct observed values. In particular, the limit does
not coarsen a training sample of 500 observations. At $N=2{,}000$ and on the
semi-synthetic tables, coordinates with more than $1{,}024$ distinct values
are binned. Binning restricts the available thresholds, while the counts at
each retained threshold are still computed from the actual observations.

\subsection{Evaluating numeric candidates efficiently}

When every coordinate is numeric and binned, a leaf stores histograms for
the joint, $X$-marginal and $T$-marginal collections.
For an $X$-split, a cumulative scan of the joint and
$X$-marginal histograms gives $n_{XT}(A_\ell)$ and $n_X(A_\ell)$ at each
cut; right-child counts follow by subtraction, and
$n_T(A_\ell)=n_T(A_r)=n_T(A)$. The roles reverse for a $T$-split.
Once the histograms are available, scanning a coordinate takes $O(H)$
operations.

After selecting a split, the builder partitions the row collections that
the split refines. For each such collection, it constructs the histogram
of the child with fewer row occurrences and obtains the other by
subtracting from the parent histogram when that histogram is retained.
The unaffected marginal collection and its histograms are shared by the
two children. Histograms can be released to bound memory use and rebuilt
from row indices when needed.

If any coordinate is categorical or has no bin limit, the builder
instead retains sorted row-index lists for the three collections. Numeric
candidates still use the histogram boundaries when a bin limit is set.
Prefix sums and a forward-moving cursor through the sorted joint rows
evaluate these candidates in a single pass. Including candidate generation,
the work is linear in the relevant row-list lengths, the number of bins
when binning is enabled, and the number of candidate cuts. Children inherit
the sorted order, so re-sorting by coordinate value is unnecessary.

\subsection{Categorical coordinates and treatment classes}

Categorical covariates and treatment classes use subset-membership splits.
For an $X$ coordinate $j$, let $\mathcal C_A$ contain the
categories present in $\mathcal I_X(A)$. For each $c\in\mathcal C_A$,
compute
\begin{equation}
  a_c=\sum_{i\in\mathcal I_{XT}(A)}\1\{X_{ij}=c\},\qquad
  b_c=\sum_{i\in\mathcal I_X(A)}\1\{X_{ij}=c\},\qquad
  s_c=\frac{a_c}{b_c}.
\end{equation}
For a treatment coordinate, replace $\mathcal I_X(A)$ by
$\mathcal I_T(A)$ and $X_{ij}$ by the treatment coordinate. Since
$\widehat p_{A_{j,c}}/\widehat q_{A_{j,c}}=N_0 s_c/n_T(A)$ for
$A_{j,c}=A\cap\{X_j=c\}$, ordering by $s_c$ is equivalent to ordering by the
local joint-to-product-marginal ratio.

For gain-based search, sort the $C=|\mathcal C_A|$ categories in increasing
order of $s_c$, breaking score ties by their first occurrence in the full
training data. The category order is recomputed at each node; first-occurrence
indices are cached and shared across forest members. For the ordered levels
$(c_1,\ldots,c_C)$, the candidate left subsets are
\begin{equation}
  \{c_1\},\ \{c_1,c_2\},\ \ldots,\ \{c_1,\ldots,c_{C-1}\}.
\end{equation}
The complement goes right. The same ordering is used for both gain
objectives. A binary treatment supplies one nontrivial class separation;
a coordinate with $C$ active categories supplies $C-1$ subsets before
validity checks. The numeric bin limits do not merge categories.

The implementation aggregates category masses to determine the ordering,
then explicitly routes the joint and relevant marginal row collections
for each candidate subset. Consequently, its candidate evaluation can
require $O(Cm)$ work for $m$ relevant row occurrences, in addition to
$O(C\log C)$ category sorting.

\subsection{Selection and exploration}

The builder selects the largest positive gain among the eligible
coordinates and leaves. Ties are resolved deterministically using node
creation order, coordinate order and a canonical ordering of the child
row-index collections. Gains within a floating-point tolerance of zero
are treated as zero. A split whose positive gain is below the minimum gain
consumes one unit of the exploration budget $k_e$. If no positive-gain
candidate exists, the builder visits pending leaves in creation order and
seeks a balanced admissible split, minimizing the difference in the children's
joint row counts. Coordinates are visited cyclically in an interleaved
$X,T$ order, and after a split the children's search starts on the opposite
side. Categorical exploration orders levels by first occurrence instead
of their ratio scores. Each such exploration split also consumes one budget
unit. Thus exploration can expose interactions after a zero-gain root split,
while positive-gain splits take priority whenever available. Growth stops
at the leaf limit or when no permitted split remains under the count,
gain and exploration constraints.

\section{Experimental setup}\label{app:experimental_protocol}

\paragraph{Response evaluation.}
We approximate the population response by averaging the known structural
outcome function over the evaluation covariates. For synthetic datasets, this
uses the independent evaluation sample of 10,000 observations; for
semi-synthetic datasets, it uses the held-out fold. Continuous-treatment grids
contain 100 equally spaced points: between the empirical 1st and 99th treatment
percentiles of the evaluation sample for Kang--Schafer, between the 2.5th and
97.5th percentiles of the full generated sample for semi-synthetic datasets,
and over $[0,1]$, including both endpoints, for Beta-mixture and SyntheticVCNet.
RMSE weights all grid points equally.

\subsection{Weight estimators}\label{app:weight-estimators}

\begin{itemize}
  \item \textbf{Permutation Weighting with Gradient Boosting (PW-GB)}
  \citep{arbour2021permutation}. We fit a histogram gradient-boosting
  classifier to distinguish the observed pairs $(X,T)$ from pairs obtained by
  permuting $T$. The classifier receives the original variables and the
  covariate--treatment interactions. We average the class probabilities over
  100 permutations before computing the observed-to-permuted odds, whose
  reciprocal is $\widehat w$.
  \item \textbf{Conditional probability model (Cond.)}. We model $P(T\mid X)$
  with histogram gradient boosting, the same learner as PW-GB. For binary
  treatment, a boosted classifier returns $P(T\mid X)$ directly. For
  continuous treatment, we fit a Gaussian conditional density with
  \texttt{skpro} \citep{gressmann2018probabilistic}: one boosted regressor predicts
  the conditional mean, and the variance is a single constant, the mean of
  the squared residuals computed out of fold by three-fold cross-fitting.
  We estimate $P(T)$ by integrating the conditional model
  over the empirical covariate distribution.
  \item \textbf{CBPS} \citep{imai2014covariate,fong2018covariate}. For categorical
  treatment, we fit a multinomial-logit propensity model by two-step GMM,
  combining likelihood-score moments with inverse-probability-weighted
  covariate-balance moments. For continuous treatment, we use a Gaussian
  treatment model and analogous GMM conditions that drive the weighted
  covariance between treatment and each covariate to zero. Both estimators use
  the overidentified specification and return the stabilized ratio
  $p(t\mid x)/p(t)$, whose reciprocal is $\widehat w$. For continuous
  treatment, the fit falls back to the exactly identified specification, which
  keeps only the balance conditions, when the GMM weighting matrix is not
  positive definite. Our implementation follows the routines in the R
  \texttt{CBPS} package \citep{fong2019package}.
\end{itemize}

\subsection{Hyperparameter selection}\label{app:search-spaces}

For a validation fold $\mathcal I_k$ of size $n_k$ and a derangement $\sigma$ of
  its indices, define the permutation log loss as
  \begin{equation}
    \label{eq:perm-logloss}
    \widehat L_k(\widehat w)
    =\frac{1}{2n_k}\sum_{i\in\mathcal I_k}\Bigl[
      \log\bigl(1+\widehat w(X_i,T_i)\bigr)
      +\log\bigl(1+\widehat w(X_i,T_{\sigma(i)})^{-1}\bigr)\Bigr].
  \end{equation}
  This is the logistic loss of a classifier that separates observed pairs
  from pairs with permuted treatment using the score $\log\widehat w$. We
  draw the permutation once per fold and use it for every candidate. All
  four partition
  estimators are selected by this common criterion, irrespective of the
  objective used to grow their trees.

We optimize every tunable weight estimator and the outcome model using 25
Optuna \citep{akiba2019optuna} trials with three-fold cross-validation and a
fixed search seed. CBPS has no tuning parameters. Weight estimators minimize the
permutation log loss of \cref{eq:perm-logloss}, averaged over the three
validation folds. BPT and BPF grow trees using the mutual-information
objective, whereas BPT$_{L^2}$ and BPF$_{L^2}$ use the squared-error
objective. All four variants are selected using \cref{eq:perm-logloss}.

For the partition estimators, we sample the marginal floor factor $\kappa$ and
the joint count floor independently on log scales, with $\kappa\in[0.25,20]$
and the joint floor in $[1,50]$. Each child must keep at least
$\kappa\sqrt{N}$ observations in each of its two marginal projections and at
least the joint floor in the cell. We fix the exploration budget to $k_e=2$,
the maximum number of leaves to 10,000, the minimum gain to $10^{-6}$ and the
bin limit to $H=1{,}024$. Each forest combines 50 trees, each grown on a
bootstrap resample of the training sample with every coordinate eligible at
every node.

For PW-GB, we sample the number of boosting iterations log-uniformly from 100
to 1,000, the maximum depth from $\{2,4,6,8\}$ and the learning rate
log-uniformly from $[0.01,0.3]$. Each candidate is evaluated using one
permutation; the selected model is refitted on 100 permutations, and the resulting
class probabilities are averaged before forming the density ratio.

Both conditional models share the boosting search space of PW-GB: we sample the number
of boosting iterations log-uniformly from 100 to 1,000, the maximum depth from
$\{2,4,6,8\}$ and the learning rate log-uniformly from $[0.01,0.3]$. The
continuous model applies the sampled configuration to its mean regressor.
Both conditional models approximate the marginal treatment
density using at most 1,000 empirical covariate observations.

The outcome model is a histogram gradient-boosting regressor selected by mean
cross-validated $R^2$. We sample its learning rate log-uniformly from
$[0.01,0.3]$, maximum number of leaves log-uniformly from $[7,127]$, minimum
leaf size log-uniformly from $[5,100]$ and $L^2$ penalty log-uniformly from
$[10^{-8},10]$.

\subsection{Data-generating processes}\label{app:dgp}

We describe the data-generating processes used in the experiments. Throughout,
$\operatorname{expit}(u)=(1+\exp(-u))^{-1}$, and random draws are mutually
independent unless stated otherwise.

\paragraph{Kang--Schafer \citep{10.1214/07-STS227}.}
Let $X=(X_1,\ldots,X_4)$, with $X_j\stackrel{\mathrm{iid}}{\sim}\mathcal
N(0,1)$, and define
\begin{align}
  \eta(X) &= X_1-0.5X_2+0.25X_3+0.1X_4, \\
  \beta(X) &= 210+27.4X_1+13.7X_2+13.7X_3+13.7X_4.
\end{align}
For binary treatment, we draw
\begin{equation}
  T\mid X\sim\operatorname{Bernoulli}\!\left(\operatorname{expit}(\eta(X))\right),
  \qquad
  Y=\beta(X)+T+\varepsilon_Y,
  \qquad \varepsilon_Y\sim\mathcal N(0,1).
\end{equation}
For continuous treatment, we extend the original
design and instead use
\begin{equation}
  T=\eta(X)+\varepsilon_T,
  \qquad
  Y=\beta(X)+\operatorname{expit}(T)+\varepsilon_Y,
  \qquad \varepsilon_T,\varepsilon_Y\sim\mathcal N(0,1).
\end{equation}
The correctly specified variants expose $X$. The misspecified variants expose
only the nonlinear transformations
\begin{align}
  Z_1 &= \exp(X_1/2),
  &Z_2 &= \frac{X_2}{1+\exp(X_1)}+10, \\
  Z_3 &= \left(\frac{X_1X_3}{25}+0.6\right)^3,
  &Z_4 &= (X_2+X_4+20)^2,
\end{align}
while treatment and outcome generation continue to depend on the latent $X$.

\paragraph{Beta-mixture.}
We designed this process to pair a treatment whose conditional distribution
is a non-Gaussian Beta mixture with a nonlinear response curve.
We draw $X\sim\mathcal N(0,I_6)$ and, once per seed, a projection
$v\sim\mathcal N(0,I_6)$. Let
\begin{equation}
  s(X)=\operatorname{expit}\!\left(\frac{X^\top v}{\sqrt{6}}\right),
  \qquad
  \widetilde s(X)=\min\!\left\{1-10^{-6},
  \max\{10^{-6},s(X)\}\right\}.
\end{equation}
The continuous-treatment design draws $\zeta\sim\operatorname{Bernoulli}(1/2)$ and
\begin{align}
  T_1\mid X &\sim
  \operatorname{Beta}\!\left(\widetilde s(X),1-\widetilde s(X)\right), \\
  T_2\mid X &\sim
  \operatorname{Beta}\!\left(99\widetilde s(X),100-99\widetilde s(X)\right),
  \qquad
  T=\zeta T_1+(1-\zeta)T_2.
\end{align}
The binary design uses
$T\mid X\sim\operatorname{Bernoulli}(\widetilde s(X))$. In both cases, the
response surface and observed outcome are
\begin{equation}
  m(X,T)=-5s(X)+5(T-0.2)^2-T^3,
  \qquad
  Y=m(X,T)+\varepsilon_Y,
  \qquad \varepsilon_Y\sim\mathcal N(0,0.5).
\end{equation}

\paragraph{SyntheticVCNet \citep{nie2021vcnet}.}
We draw $X_j\stackrel{\mathrm{iid}}{\sim}\operatorname{Unif}(0,1)$ for
$j=1,\ldots,6$. Define
\begin{align}
  h(X)={}&
  \frac{10\sin(\max\{X_1,X_2,X_3\})+\max\{X_3,X_4,X_5\}^3}
       {1+(X_1+X_5)^2} \nonumber\\
  &+\sin(0.5X_3)\left\{1+\exp(X_4-0.5X_3)\right\}
  +X_3^2+2\sin(X_4)+2X_5-6.5.
\end{align}
Treatment and outcome are generated as
\begin{align}
  T &= \operatorname{expit}\!\left(h(X)+\varepsilon_T\right),
  &\varepsilon_T&\sim\mathcal N(0,0.25), \\
  m(X,T) &= \cos\!\left(2\pi(T-0.5)\right)
  \left\{T^2+
  \frac{4\max\{X_1,X_6\}^3}{1+2X_3^2}\sin(X_4)\right\}, \nonumber\\
  Y &= m(X,T)+\varepsilon_Y,
  &\varepsilon_Y&\sim\mathcal N(0,0.25).
\end{align}

\paragraph{Semi-synthetic datasets.}
The Digits \citep{digits}, ACIC~2016 \citep{dorie2019automated, niswander1972women} and California housing \citep{pace1997sparse} settings share the same mechanism
and differ only in their empirical covariate distribution. Inspired by ACIC~2016 \citep{dorie2019automated}, we
define the following semi-synthetic data-generating process. Let $X_1,\ldots,X_N$
denote the column-standardized covariates from one of these tables. For each
varying column $j$, we transform its average rank $R_{ij}$ to a normal score,
\begin{equation}
  Z_{ij}=\frac{\Phi^{-1}\!\left((R_{ij}-1/2)/N\right)}{s_j},
\end{equation}
where $s_j$ is the empirical standard deviation of the numerator. We then
select $|\mathcal S|=\min\{8,\lfloor p_v/2\rfloor\}$ columns, where $p_v$ is
the number of varying columns. For every selected column, we sample a
coefficient $a_j\sim\mathcal N(0,1)$ and choose a transform $\phi_j(Z_j)$ uniformly from
\begin{equation}
  Z_j,\qquad Z_j^2,\qquad
  \1\{Z_j>\widehat F_j^{-1}(U_j)\},\qquad
  U_j\sim\operatorname{Unif}(0.2,0.8),
\end{equation}
where $\widehat F_j^{-1}$ is the empirical quantile function. We also draw
$\lfloor|\mathcal S|/2\rfloor$ pairwise products between selected columns.
Hence the unstandardized assignment function has the form
\begin{equation}
  \widetilde\psi(Z)=\sum_{j\in\mathcal S}a_j\phi_j(Z_j)
  +\sum_{\ell=1}^{\lfloor|\mathcal S|/2\rfloor}
  d_\ell Z_{j_\ell}Z_{k_\ell},
  \qquad d_\ell\sim\mathcal N(0,1),
\end{equation}
and we set $\psi(Z)=(\widetilde\psi(Z)-\overline\psi)/s_\psi$, where $\overline\psi$ and
$s_\psi$ are its empirical mean and standard deviation over the table. Treatment
is generated by
\begin{equation}
  T=\psi(Z)+1.5\varepsilon_T,
  \qquad \varepsilon_T\sim\mathcal N(0,1).
\end{equation}
To generate the treatment effect, let $\mathbf B(T)\in\mathbb R^K$ be a cubic B-spline
basis with six knots fitted over the observed treatment range, and draw
$\theta\sim\mathcal N(0,I_K/K)$. We center $\mathbf B(t)^\top\theta$ over the observed treatments
and scale it to unit empirical standard deviation, obtaining $\gamma(t)$. Finally,
\begin{equation}
  Y=2\psi(Z)+\gamma(T)+\varepsilon_Y,
  \qquad \varepsilon_Y\sim\mathcal N(0,1).
\end{equation}
A new assignment function and treatment-effect curve are drawn for each seed.

\clearpage
\section{Forest aggregation}\label{app:forest-ablation}

We compare harmonic, geometric and arithmetic means of the tree weights.
Each variant retains 50 trees
and the main experiment's datasets, seeds/folds and count-floor search space,
with independent 25-trial permutation-log-loss tuning.

\begin{table}[htbp]
\centering
\small
\setlength{\tabcolsep}{6pt}
\begin{adjustbox}{max width=\linewidth}
\begin{tabular}{lrrrr}
\toprule
Forest configuration & Weight estimation & \multicolumn{3}{c}{Response RMSE} \\
\cmidrule(lr){1-1}\cmidrule(lr){2-2}\cmidrule(lr){3-5}
Weight mean & Perm. log loss & MSM & GPS & DR \\
\midrule
\multicolumn{5}{l}{\textit{MI gain (BPF)}} \\
Harmonic & \textbf{1.00} & \textbf{1.53} & \textbf{1.47} & \textbf{1.59} \\
Geometric & 2.06 & 1.94 & 2.06 & 1.94 \\
Arithmetic & 2.94 & 2.53 & 2.47 & 2.47 \\
\midrule
\multicolumn{5}{l}{\textit{$L^2$ gain (BPF$_{L^2}$)}} \\
Harmonic & \textbf{1.00} & \textbf{1.71} & \textbf{1.53} & \textbf{1.53} \\
Geometric & 2.00 & 1.76 & 1.94 & 1.76 \\
Arithmetic & 3.00 & 2.53 & 2.53 & 2.71 \\
\bottomrule
\end{tabular}
\end{adjustbox}
\caption{Forest aggregation ablation. Entries are
mean ranks (lower is better) of held-out permutation log loss and response
RMSE. We rank the unrounded five-seed/fold means among the three variants
within each split objective, using midranks for ties, then average equally
over the 17 dataset--sample-size settings. Bold marks the lowest rank in each
panel and column. All variants are independently tuned.}
\label{tab:forest_ablation}
\end{table}

Harmonic averaging has the lowest held-out
permutation log loss in all 17 settings under both split objectives.

\clearpage
\section{Extended experimental results}\label{app:extended-results}

This appendix collects the full result tables and figures behind
\cref{sec:experiments}: GPS and DR causal-response errors, grouped ranks, weight
stability, computation times and the estimated dose-response curves.

\subsection{Causal-response results}\label{app:response-results}

\Cref{tab:causal_msm_response,tab:causal_gps_response,tab:causal_dr_response}
report the response RMSE of the MSM, GPS and DR estimators, respectively, for
every weight estimator, dataset and training size.

\begin{table}[htbp]
\centering
\footnotesize
\setlength{\tabcolsep}{4pt}
\begin{adjustbox}{max width=\linewidth}
\begin{tabular}{lrrrrrrr}
\toprule
Dataset & BPT & BPF & BPT$_{L^2}$ & BPF$_{L^2}$ & PW-GB & CBPS & Cond. \\
\midrule
\multicolumn{8}{l}{\textit{Synthetic, $N=500$}} \\
KS binary & \makecell[r]{${8.156}^{(6)}$\\[-1.5pt]$\pm 3.412$} & \makecell[r]{${2.740}^{(2)}$\\[-1.5pt]$\pm 1.148$} & \makecell[r]{${9.433}^{(7)}$\\[-1.5pt]$\pm 5.371$} & \makecell[r]{${3.216}^{(4)}$\\[-1.5pt]$\pm 1.065$} & \makecell[r]{${\mathbf{1.895}}^{(1)}$\\[-1.5pt]$\pm 1.365$} & \makecell[r]{${2.939}^{(3)}$\\[-1.5pt]$\pm 1.163$} & \makecell[r]{${6.003}^{(5)}$\\[-1.5pt]$\pm 1.583$} \\
KS binary, misspec. & \makecell[r]{${7.692}^{(7)}$\\[-1.5pt]$\pm 2.079$} & \makecell[r]{${3.120}^{(2)}$\\[-1.5pt]$\pm 0.759$} & \makecell[r]{${5.799}^{(6)}$\\[-1.5pt]$\pm 1.819$} & \makecell[r]{${\mathbf{2.458}}^{(1)}$\\[-1.5pt]$\pm 1.309$} & \makecell[r]{${3.548}^{(3)}$\\[-1.5pt]$\pm 1.643$} & \makecell[r]{${4.775}^{(4)}$\\[-1.5pt]$\pm 1.412$} & \makecell[r]{${5.733}^{(5)}$\\[-1.5pt]$\pm 2.800$} \\
Beta-mixture binary & \makecell[r]{${0.298}^{(6)}$\\[-1.5pt]$\pm 0.089$} & \makecell[r]{${0.212}^{(4)}$\\[-1.5pt]$\pm 0.064$} & \makecell[r]{${0.336}^{(7)}$\\[-1.5pt]$\pm 0.114$} & \makecell[r]{${0.207}^{(3)}$\\[-1.5pt]$\pm 0.054$} & \makecell[r]{${0.160}^{(2)}$\\[-1.5pt]$\pm 0.011$} & \makecell[r]{${\mathbf{0.106}}^{(1)}$\\[-1.5pt]$\pm 0.031$} & \makecell[r]{${0.252}^{(5)}$\\[-1.5pt]$\pm 0.069$} \\
KS continuous & \makecell[r]{${10.564}^{(3)}$\\[-1.5pt]$\pm 2.568$} & \makecell[r]{${10.254}^{(2)}$\\[-1.5pt]$\pm 1.272$} & \makecell[r]{${13.594}^{(6)}$\\[-1.5pt]$\pm 3.743$} & \makecell[r]{${\mathbf{9.560}}^{(1)}$\\[-1.5pt]$\pm 0.550$} & \makecell[r]{${12.483}^{(5)}$\\[-1.5pt]$\pm 2.356$} & \makecell[r]{${11.824}^{(4)}$\\[-1.5pt]$\pm 1.507$} & \makecell[r]{${14.414}^{(7)}$\\[-1.5pt]$\pm 1.379$} \\
KS continuous, misspec. & \makecell[r]{${12.381}^{(4)}$\\[-1.5pt]$\pm 2.113$} & \makecell[r]{${\mathbf{10.118}}^{(1)}$\\[-1.5pt]$\pm 1.540$} & \makecell[r]{${12.728}^{(5)}$\\[-1.5pt]$\pm 3.035$} & \makecell[r]{${10.952}^{(2)}$\\[-1.5pt]$\pm 1.954$} & \makecell[r]{${11.625}^{(3)}$\\[-1.5pt]$\pm 2.033$} & \makecell[r]{${21.426}^{(7)}$\\[-1.5pt]$\pm 4.431$} & \makecell[r]{${17.393}^{(6)}$\\[-1.5pt]$\pm 1.946$} \\
Beta-mixture continuous & \makecell[r]{${0.618}^{(7)}$\\[-1.5pt]$\pm 0.155$} & \makecell[r]{${0.314}^{(2)}$\\[-1.5pt]$\pm 0.087$} & \makecell[r]{${0.574}^{(6)}$\\[-1.5pt]$\pm 0.173$} & \makecell[r]{${0.356}^{(3)}$\\[-1.5pt]$\pm 0.125$} & \makecell[r]{${0.356}^{(4)}$\\[-1.5pt]$\pm 0.090$} & \makecell[r]{${\mathbf{0.284}}^{(1)}$\\[-1.5pt]$\pm 0.044$} & \makecell[r]{${0.426}^{(5)}$\\[-1.5pt]$\pm 0.084$} \\
SyntheticVCNet & \makecell[r]{${0.209}^{(6)}$\\[-1.5pt]$\pm 0.038$} & \makecell[r]{${\mathbf{0.188}}^{(1)}$\\[-1.5pt]$\pm 0.033$} & \makecell[r]{${0.206}^{(4)}$\\[-1.5pt]$\pm 0.047$} & \makecell[r]{${0.202}^{(3)}$\\[-1.5pt]$\pm 0.032$} & \makecell[r]{${0.200}^{(2)}$\\[-1.5pt]$\pm 0.044$} & \makecell[r]{${0.207}^{(5)}$\\[-1.5pt]$\pm 0.052$} & \makecell[r]{${0.213}^{(7)}$\\[-1.5pt]$\pm 0.037$} \\
\addlinespace[3pt]
Mean rank & 5.57 & 2.00 & 5.86 & 2.43 & 2.86 & 3.57 & 5.71 \\
\midrule
\multicolumn{8}{l}{\textit{Synthetic, $N=2{,}000$}} \\
KS binary & \makecell[r]{${5.183}^{(6)}$\\[-1.5pt]$\pm 1.729$} & \makecell[r]{${2.053}^{(3)}$\\[-1.5pt]$\pm 0.836$} & \makecell[r]{${8.435}^{(7)}$\\[-1.5pt]$\pm 4.130$} & \makecell[r]{${2.275}^{(4)}$\\[-1.5pt]$\pm 0.539$} & \makecell[r]{${1.847}^{(2)}$\\[-1.5pt]$\pm 0.669$} & \makecell[r]{${\mathbf{1.642}}^{(1)}$\\[-1.5pt]$\pm 1.259$} & \makecell[r]{${4.692}^{(5)}$\\[-1.5pt]$\pm 0.420$} \\
KS binary, misspec. & \makecell[r]{${5.462}^{(6)}$\\[-1.5pt]$\pm 2.055$} & \makecell[r]{${2.865}^{(2)}$\\[-1.5pt]$\pm 0.906$} & \makecell[r]{${8.049}^{(7)}$\\[-1.5pt]$\pm 3.908$} & \makecell[r]{${\mathbf{2.682}}^{(1)}$\\[-1.5pt]$\pm 1.046$} & \makecell[r]{${3.021}^{(3)}$\\[-1.5pt]$\pm 0.611$} & \makecell[r]{${4.462}^{(4)}$\\[-1.5pt]$\pm 0.501$} & \makecell[r]{${4.862}^{(5)}$\\[-1.5pt]$\pm 0.798$} \\
Beta-mixture binary & \makecell[r]{${0.253}^{(6)}$\\[-1.5pt]$\pm 0.123$} & \makecell[r]{${0.130}^{(3)}$\\[-1.5pt]$\pm 0.058$} & \makecell[r]{${0.264}^{(7)}$\\[-1.5pt]$\pm 0.119$} & \makecell[r]{${0.140}^{(4)}$\\[-1.5pt]$\pm 0.065$} & \makecell[r]{${0.095}^{(2)}$\\[-1.5pt]$\pm 0.049$} & \makecell[r]{${\mathbf{0.067}}^{(1)}$\\[-1.5pt]$\pm 0.038$} & \makecell[r]{${0.212}^{(5)}$\\[-1.5pt]$\pm 0.100$} \\
KS continuous & \makecell[r]{${8.271}^{(5)}$\\[-1.5pt]$\pm 1.665$} & \makecell[r]{${\mathbf{6.600}}^{(1)}$\\[-1.5pt]$\pm 1.311$} & \makecell[r]{${9.212}^{(6)}$\\[-1.5pt]$\pm 0.998$} & \makecell[r]{${6.953}^{(2)}$\\[-1.5pt]$\pm 1.000$} & \makecell[r]{${7.722}^{(3)}$\\[-1.5pt]$\pm 1.906$} & \makecell[r]{${8.033}^{(4)}$\\[-1.5pt]$\pm 1.412$} & \makecell[r]{${12.490}^{(7)}$\\[-1.5pt]$\pm 1.758$} \\
KS continuous, misspec. & \makecell[r]{${10.080}^{(5)}$\\[-1.5pt]$\pm 2.929$} & \makecell[r]{${7.417}^{(2)}$\\[-1.5pt]$\pm 1.192$} & \makecell[r]{${9.808}^{(4)}$\\[-1.5pt]$\pm 2.062$} & \makecell[r]{${\mathbf{6.883}}^{(1)}$\\[-1.5pt]$\pm 1.168$} & \makecell[r]{${9.779}^{(3)}$\\[-1.5pt]$\pm 2.053$} & \makecell[r]{${21.464}^{(7)}$\\[-1.5pt]$\pm 2.805$} & \makecell[r]{${11.171}^{(6)}$\\[-1.5pt]$\pm 1.695$} \\
Beta-mixture continuous & \makecell[r]{${0.369}^{(6)}$\\[-1.5pt]$\pm 0.134$} & \makecell[r]{${0.207}^{(2)}$\\[-1.5pt]$\pm 0.064$} & \makecell[r]{${0.391}^{(7)}$\\[-1.5pt]$\pm 0.159$} & \makecell[r]{${0.243}^{(4)}$\\[-1.5pt]$\pm 0.101$} & \makecell[r]{${\mathbf{0.200}}^{(1)}$\\[-1.5pt]$\pm 0.040$} & \makecell[r]{${0.208}^{(3)}$\\[-1.5pt]$\pm 0.047$} & \makecell[r]{${0.300}^{(5)}$\\[-1.5pt]$\pm 0.042$} \\
SyntheticVCNet & \makecell[r]{${0.162}^{(4)}$\\[-1.5pt]$\pm 0.014$} & \makecell[r]{${0.156}^{(3)}$\\[-1.5pt]$\pm 0.029$} & \makecell[r]{${\mathbf{0.141}}^{(1)}$\\[-1.5pt]$\pm 0.011$} & \makecell[r]{${0.146}^{(2)}$\\[-1.5pt]$\pm 0.022$} & \makecell[r]{${0.163}^{(5)}$\\[-1.5pt]$\pm 0.022$} & \makecell[r]{${0.239}^{(7)}$\\[-1.5pt]$\pm 0.027$} & \makecell[r]{${0.178}^{(6)}$\\[-1.5pt]$\pm 0.034$} \\
\addlinespace[3pt]
Mean rank & 5.43 & 2.29 & 5.57 & 2.57 & 2.71 & 3.86 & 5.57 \\
\midrule
\multicolumn{8}{l}{\textit{Semi-synthetic}} \\
Digits & \makecell[r]{${0.924}^{(7)}$\\[-1.5pt]$\pm 0.366$} & \makecell[r]{${\mathbf{0.632}}^{(1)}$\\[-1.5pt]$\pm 0.261$} & \makecell[r]{${0.844}^{(5)}$\\[-1.5pt]$\pm 0.346$} & \makecell[r]{${0.689}^{(2)}$\\[-1.5pt]$\pm 0.280$} & \makecell[r]{${0.714}^{(3)}$\\[-1.5pt]$\pm 0.294$} & \makecell[r]{${0.867}^{(6)}$\\[-1.5pt]$\pm 0.354$} & \makecell[r]{${0.758}^{(4)}$\\[-1.5pt]$\pm 0.298$} \\
ACIC & \makecell[r]{${0.713}^{(6)}$\\[-1.5pt]$\pm 0.171$} & \makecell[r]{${0.427}^{(3)}$\\[-1.5pt]$\pm 0.120$} & \makecell[r]{${0.734}^{(7)}$\\[-1.5pt]$\pm 0.203$} & \makecell[r]{${0.420}^{(2)}$\\[-1.5pt]$\pm 0.108$} & \makecell[r]{${\mathbf{0.351}}^{(1)}$\\[-1.5pt]$\pm 0.046$} & \makecell[r]{${0.433}^{(4)}$\\[-1.5pt]$\pm 0.115$} & \makecell[r]{${0.530}^{(5)}$\\[-1.5pt]$\pm 0.125$} \\
California & \makecell[r]{${0.504}^{(6)}$\\[-1.5pt]$\pm 0.079$} & \makecell[r]{${0.275}^{(2)}$\\[-1.5pt]$\pm 0.060$} & \makecell[r]{${0.499}^{(5)}$\\[-1.5pt]$\pm 0.116$} & \makecell[r]{${0.281}^{(3)}$\\[-1.5pt]$\pm 0.048$} & \makecell[r]{${\mathbf{0.254}}^{(1)}$\\[-1.5pt]$\pm 0.057$} & \makecell[r]{${0.904}^{(7)}$\\[-1.5pt]$\pm 0.267$} & \makecell[r]{${0.326}^{(4)}$\\[-1.5pt]$\pm 0.029$} \\
\addlinespace[3pt]
Mean rank & 6.33 & 2.00 & 5.67 & 2.33 & 1.67 & 5.67 & 4.33 \\
\midrule
Overall mean rank & 5.65 & 2.12 & 5.71 & 2.47 & 2.59 & 4.06 & 5.41 \\
\bottomrule
\end{tabular}
\end{adjustbox}
\caption{GPS response RMSE. Mean $\pm$ sample SD over five seeds/folds. Parenthesized ranks
use unrounded means per dataset, with midranks for ties and bold
for the smallest mean.
BPT and BPF grow trees with the mutual-information gain and
the $L^2$ variants with the squared-error gain. Mean rank averages the ranks of each panel. Overall mean
rank gives equal weight to each dataset--sample-size combination. Lower ranks
are better.}
\label{tab:causal_gps_response}
\end{table}

\begin{table}[htbp]
\centering
\footnotesize
\setlength{\tabcolsep}{4pt}
\begin{adjustbox}{max width=\linewidth}
\begin{tabular}{lrrrrrrr}
\toprule
Dataset & BPT & BPF & BPT$_{L^2}$ & BPF$_{L^2}$ & PW-GB & CBPS & Cond. \\
\midrule
\multicolumn{8}{l}{\textit{Synthetic, $N=500$}} \\
KS binary & \makecell[r]{${1.411}^{(4)}$\\[-1.5pt]$\pm 1.514$} & \makecell[r]{${\mathbf{1.315}}^{(1)}$\\[-1.5pt]$\pm 1.499$} & \makecell[r]{${1.434}^{(5)}$\\[-1.5pt]$\pm 1.499$} & \makecell[r]{${1.345}^{(2)}$\\[-1.5pt]$\pm 1.467$} & \makecell[r]{${1.404}^{(3)}$\\[-1.5pt]$\pm 1.361$} & \makecell[r]{${1.556}^{(6)}$\\[-1.5pt]$\pm 1.367$} & \makecell[r]{${35.468}^{(7)}$\\[-1.5pt]$\pm 28.141$} \\
KS binary, misspec. & \makecell[r]{${\mathbf{1.757}}^{(1)}$\\[-1.5pt]$\pm 1.420$} & \makecell[r]{${1.807}^{(2)}$\\[-1.5pt]$\pm 1.469$} & \makecell[r]{${1.908}^{(4)}$\\[-1.5pt]$\pm 1.385$} & \makecell[r]{${1.945}^{(5)}$\\[-1.5pt]$\pm 1.267$} & \makecell[r]{${1.850}^{(3)}$\\[-1.5pt]$\pm 1.221$} & \makecell[r]{${2.496}^{(6)}$\\[-1.5pt]$\pm 1.231$} & \makecell[r]{${175.097}^{(7)}$\\[-1.5pt]$\pm 134.516$} \\
Beta-mixture binary & \makecell[r]{${0.145}^{(5)}$\\[-1.5pt]$\pm 0.036$} & \makecell[r]{${0.105}^{(2)}$\\[-1.5pt]$\pm 0.033$} & \makecell[r]{${0.147}^{(6)}$\\[-1.5pt]$\pm 0.034$} & \makecell[r]{${0.112}^{(4)}$\\[-1.5pt]$\pm 0.033$} & \makecell[r]{${0.110}^{(3)}$\\[-1.5pt]$\pm 0.018$} & \makecell[r]{${\mathbf{0.053}}^{(1)}$\\[-1.5pt]$\pm 0.029$} & \makecell[r]{${18.421}^{(7)}$\\[-1.5pt]$\pm 25.641$} \\
KS continuous & \makecell[r]{${2.570}^{(2)}$\\[-1.5pt]$\pm 1.144$} & \makecell[r]{${\mathbf{2.466}}^{(1)}$\\[-1.5pt]$\pm 1.208$} & \makecell[r]{${2.630}^{(4)}$\\[-1.5pt]$\pm 1.181$} & \makecell[r]{${2.601}^{(3)}$\\[-1.5pt]$\pm 1.097$} & \makecell[r]{${2.905}^{(5)}$\\[-1.5pt]$\pm 1.265$} & \makecell[r]{${3.856}^{(6)}$\\[-1.5pt]$\pm 0.829$} & \makecell[r]{${5.095}^{(7)}$\\[-1.5pt]$\pm 2.767$} \\
KS continuous, misspec. & \makecell[r]{${\mathbf{4.286}}^{(1)}$\\[-1.5pt]$\pm 0.662$} & \makecell[r]{${4.564}^{(2)}$\\[-1.5pt]$\pm 0.511$} & \makecell[r]{${4.570}^{(3)}$\\[-1.5pt]$\pm 0.622$} & \makecell[r]{${4.810}^{(4)}$\\[-1.5pt]$\pm 0.410$} & \makecell[r]{${4.866}^{(5)}$\\[-1.5pt]$\pm 0.694$} & \makecell[r]{${1.57\mathrm{e}{+4}}^{(7)}$\\[-1.5pt]$\pm 3.11\mathrm{e}{+4}$} & \makecell[r]{${37.433}^{(6)}$\\[-1.5pt]$\pm 65.443$} \\
Beta-mixture continuous & \makecell[r]{${0.232}^{(5)}$\\[-1.5pt]$\pm 0.054$} & \makecell[r]{${0.184}^{(2)}$\\[-1.5pt]$\pm 0.066$} & \makecell[r]{${0.209}^{(4)}$\\[-1.5pt]$\pm 0.078$} & \makecell[r]{${0.188}^{(3)}$\\[-1.5pt]$\pm 0.072$} & \makecell[r]{${\mathbf{0.183}}^{(1)}$\\[-1.5pt]$\pm 0.050$} & \makecell[r]{${0.649}^{(6)}$\\[-1.5pt]$\pm 0.475$} & \makecell[r]{${1.151}^{(7)}$\\[-1.5pt]$\pm 1.241$} \\
SyntheticVCNet & \makecell[r]{${0.160}^{(5)}$\\[-1.5pt]$\pm 0.062$} & \makecell[r]{${0.148}^{(3)}$\\[-1.5pt]$\pm 0.042$} & \makecell[r]{${\mathbf{0.143}}^{(1)}$\\[-1.5pt]$\pm 0.044$} & \makecell[r]{${0.146}^{(2)}$\\[-1.5pt]$\pm 0.046$} & \makecell[r]{${0.151}^{(4)}$\\[-1.5pt]$\pm 0.036$} & \makecell[r]{${0.264}^{(6)}$\\[-1.5pt]$\pm 0.127$} & \makecell[r]{${1.453}^{(7)}$\\[-1.5pt]$\pm 2.194$} \\
\addlinespace[3pt]
Mean rank & 3.29 & 1.86 & 3.86 & 3.29 & 3.43 & 5.43 & 6.86 \\
\midrule
\multicolumn{8}{l}{\textit{Synthetic, $N=2{,}000$}} \\
KS binary & \makecell[r]{${0.700}^{(5)}$\\[-1.5pt]$\pm 0.690$} & \makecell[r]{${0.676}^{(3)}$\\[-1.5pt]$\pm 0.722$} & \makecell[r]{${0.691}^{(4)}$\\[-1.5pt]$\pm 0.721$} & \makecell[r]{${0.676}^{(2)}$\\[-1.5pt]$\pm 0.736$} & \makecell[r]{${\mathbf{0.640}}^{(1)}$\\[-1.5pt]$\pm 0.753$} & \makecell[r]{${0.723}^{(6)}$\\[-1.5pt]$\pm 0.654$} & \makecell[r]{${40.776}^{(7)}$\\[-1.5pt]$\pm 59.821$} \\
KS binary, misspec. & \makecell[r]{${\mathbf{0.824}}^{(1)}$\\[-1.5pt]$\pm 0.538$} & \makecell[r]{${0.836}^{(2)}$\\[-1.5pt]$\pm 0.560$} & \makecell[r]{${0.859}^{(4)}$\\[-1.5pt]$\pm 0.555$} & \makecell[r]{${0.841}^{(3)}$\\[-1.5pt]$\pm 0.581$} & \makecell[r]{${0.886}^{(5)}$\\[-1.5pt]$\pm 0.552$} & \makecell[r]{${1.118}^{(6)}$\\[-1.5pt]$\pm 0.772$} & \makecell[r]{${63.458}^{(7)}$\\[-1.5pt]$\pm 75.931$} \\
Beta-mixture binary & \makecell[r]{${0.049}^{(3)}$\\[-1.5pt]$\pm 0.027$} & \makecell[r]{${0.051}^{(5)}$\\[-1.5pt]$\pm 0.026$} & \makecell[r]{${0.056}^{(6)}$\\[-1.5pt]$\pm 0.028$} & \makecell[r]{${0.048}^{(2)}$\\[-1.5pt]$\pm 0.023$} & \makecell[r]{${0.050}^{(4)}$\\[-1.5pt]$\pm 0.022$} & \makecell[r]{${\mathbf{0.038}}^{(1)}$\\[-1.5pt]$\pm 0.022$} & \makecell[r]{${4.618}^{(7)}$\\[-1.5pt]$\pm 7.484$} \\
KS continuous & \makecell[r]{${0.977}^{(3)}$\\[-1.5pt]$\pm 0.644$} & \makecell[r]{${\mathbf{0.823}}^{(1)}$\\[-1.5pt]$\pm 0.525$} & \makecell[r]{${1.114}^{(5)}$\\[-1.5pt]$\pm 0.588$} & \makecell[r]{${0.917}^{(2)}$\\[-1.5pt]$\pm 0.490$} & \makecell[r]{${1.026}^{(4)}$\\[-1.5pt]$\pm 0.469$} & \makecell[r]{${2.011}^{(6)}$\\[-1.5pt]$\pm 0.748$} & \makecell[r]{${4.473}^{(7)}$\\[-1.5pt]$\pm 6.308$} \\
KS continuous, misspec. & \makecell[r]{${1.684}^{(3)}$\\[-1.5pt]$\pm 0.379$} & \makecell[r]{${1.553}^{(2)}$\\[-1.5pt]$\pm 0.312$} & \makecell[r]{${1.742}^{(4)}$\\[-1.5pt]$\pm 0.401$} & \makecell[r]{${\mathbf{1.548}}^{(1)}$\\[-1.5pt]$\pm 0.341$} & \makecell[r]{${2.413}^{(5)}$\\[-1.5pt]$\pm 1.457$} & \makecell[r]{${88.121}^{(7)}$\\[-1.5pt]$\pm 187.817$} & \makecell[r]{${4.608}^{(6)}$\\[-1.5pt]$\pm 5.017$} \\
Beta-mixture continuous & \makecell[r]{${0.123}^{(5)}$\\[-1.5pt]$\pm 0.053$} & \makecell[r]{${\mathbf{0.102}}^{(1)}$\\[-1.5pt]$\pm 0.046$} & \makecell[r]{${0.122}^{(4)}$\\[-1.5pt]$\pm 0.051$} & \makecell[r]{${0.109}^{(2)}$\\[-1.5pt]$\pm 0.045$} & \makecell[r]{${0.109}^{(3)}$\\[-1.5pt]$\pm 0.022$} & \makecell[r]{${0.997}^{(7)}$\\[-1.5pt]$\pm 1.818$} & \makecell[r]{${0.323}^{(6)}$\\[-1.5pt]$\pm 0.182$} \\
SyntheticVCNet & \makecell[r]{${0.130}^{(4)}$\\[-1.5pt]$\pm 0.027$} & \makecell[r]{${0.117}^{(3)}$\\[-1.5pt]$\pm 0.017$} & \makecell[r]{${\mathbf{0.109}}^{(1)}$\\[-1.5pt]$\pm 0.027$} & \makecell[r]{${0.114}^{(2)}$\\[-1.5pt]$\pm 0.004$} & \makecell[r]{${0.136}^{(6)}$\\[-1.5pt]$\pm 0.044$} & \makecell[r]{${0.134}^{(5)}$\\[-1.5pt]$\pm 0.049$} & \makecell[r]{${0.333}^{(7)}$\\[-1.5pt]$\pm 0.197$} \\
\addlinespace[3pt]
Mean rank & 3.43 & 2.43 & 4.00 & 2.00 & 4.00 & 5.43 & 6.71 \\
\midrule
\multicolumn{8}{l}{\textit{Semi-synthetic}} \\
Digits & \makecell[r]{${0.472}^{(4)}$\\[-1.5pt]$\pm 0.166$} & \makecell[r]{${0.469}^{(2)}$\\[-1.5pt]$\pm 0.153$} & \makecell[r]{${0.473}^{(5)}$\\[-1.5pt]$\pm 0.171$} & \makecell[r]{${0.469}^{(3)}$\\[-1.5pt]$\pm 0.164$} & \makecell[r]{${\mathbf{0.459}}^{(1)}$\\[-1.5pt]$\pm 0.152$} & \makecell[r]{${2.60\mathrm{e}{+6}}^{(7)}$\\[-1.5pt]$\pm 5.42\mathrm{e}{+6}$} & \makecell[r]{${2.46\mathrm{e}{+6}}^{(6)}$\\[-1.5pt]$\pm 5.49\mathrm{e}{+6}$} \\
ACIC & \makecell[r]{${0.315}^{(3)}$\\[-1.5pt]$\pm 0.110$} & \makecell[r]{${\mathbf{0.311}}^{(1)}$\\[-1.5pt]$\pm 0.113$} & \makecell[r]{${0.319}^{(4)}$\\[-1.5pt]$\pm 0.113$} & \makecell[r]{${0.312}^{(2)}$\\[-1.5pt]$\pm 0.113$} & \makecell[r]{${0.335}^{(5)}$\\[-1.5pt]$\pm 0.175$} & \makecell[r]{${1.44\mathrm{e}{+6}}^{(6)}$\\[-1.5pt]$\pm 3.21\mathrm{e}{+6}$} & \makecell[r]{${2.23\mathrm{e}{+6}}^{(7)}$\\[-1.5pt]$\pm 3.09\mathrm{e}{+6}$} \\
California & \makecell[r]{${0.268}^{(2)}$\\[-1.5pt]$\pm 0.165$} & \makecell[r]{${0.273}^{(5)}$\\[-1.5pt]$\pm 0.166$} & \makecell[r]{${0.271}^{(3)}$\\[-1.5pt]$\pm 0.160$} & \makecell[r]{${0.273}^{(4)}$\\[-1.5pt]$\pm 0.163$} & \makecell[r]{${\mathbf{0.257}}^{(1)}$\\[-1.5pt]$\pm 0.163$} & \makecell[r]{${1.11\mathrm{e}{+5}}^{(7)}$\\[-1.5pt]$\pm 1.45\mathrm{e}{+5}$} & \makecell[r]{${0.308}^{(6)}$\\[-1.5pt]$\pm 0.143$} \\
\addlinespace[3pt]
Mean rank & 3.00 & 2.67 & 4.00 & 3.00 & 2.33 & 6.67 & 6.33 \\
\midrule
Overall mean rank & 3.29 & 2.24 & 3.94 & 2.71 & 3.47 & 5.65 & 6.71 \\
\bottomrule
\end{tabular}
\end{adjustbox}
\caption{DR response RMSE. Mean $\pm$ sample SD over five seeds/folds. Parenthesized ranks
use unrounded means per dataset, with midranks for ties and bold
for the smallest mean.
BPT and BPF grow trees with the mutual-information gain and
the $L^2$ variants with the squared-error gain. Mean rank averages the ranks of each panel. Overall mean
rank gives equal weight to each dataset--sample-size combination. Lower ranks
are better.}
\label{tab:causal_dr_response}
\end{table}

\paragraph{GPS and DR accuracy.}
Across the 17 dataset--sample-size settings, BPF has the lowest
mean RMSE ranks under GPS and DR: 2.12 and 2.24, respectively,
compared with 2.59 and 3.47 for PW-GB
(\cref{tab:causal_response_summary}).
Ranking BPF against the three baselines alone gives mean ranks of
1.65 and 1.41, compared with 1.88 for PW-GB under both estimators.
The ordering depends on the dataset: PW-GB achieves the lowest
GPS error on ACIC and California and the lowest DR error on
Digits and California. CBPS achieves the lowest GPS and DR errors
on Beta-mixture binary at both sample sizes.

\paragraph{Forests and split objectives under GPS and DR.}
Under GPS, BPF improves on BPT in all 17 settings, and
BPF$_{L^2}$ improves on BPT$_{L^2}$ in 16 of 17.
The improvement is less uniform under DR, where each forest
improves on its corresponding tree in 12 of 17 settings.
In the semi-synthetic DR experiments, the partition methods
have similar mean errors relative to their between-fold variation.
BPF has a better overall mean rank than BPF$_{L^2}$ under both
GPS and DR. Harmonic averaging also has the lowest mean rank
for GPS and DR response RMSE under both split objectives
(\cref{tab:forest_ablation}).

\paragraph{Large errors under doubly robust estimation.}
The DR results reveal pronounced instability for some baseline weights,
despite a numerical cap of $10^8$ on the weights used in the pseudo-outcome.
On Digits and ACIC, mean RMSE exceeds $10^6$ for both CBPS
and Cond., whereas the partition estimators and PW-GB remain
below 0.5.
This behavior is consistent with amplification of outcome
residuals by extreme estimated weights in
\cref{eq:dr-estimator}.
Across the 55 continuous-treatment fits of the weight-estimation stage,
CBPS produces a maximum held-out weight above $10^3$ in 17 fits
and reaches approximately $4\times10^9$.
The partition estimators remain below 60 and PW-GB below 160
(\cref{tab:causal_maximum_weights}).
These diagnostics report raw estimated weights before the numerical cap used
in DR estimation.

\subsection{Weight stability}\label{app:weight-stability}

\Cref{tab:causal_maximum_weights} reports the largest held-out weight
of each selected weight estimator on the continuous-treatment datasets, together
with the number of runs whose largest weight exceeds $10^3$.

\begin{table}[htbp]
\centering
\footnotesize
\setlength{\tabcolsep}{4pt}
\begin{adjustbox}{max width=\linewidth}
\begin{tabular}{lrrrrrrr}
\toprule
Dataset & BPT & BPF & BPT$_{L^2}$ & BPF$_{L^2}$ & PW-GB & CBPS & Cond. \\
\midrule
\multicolumn{8}{l}{\textit{Synthetic, $N=500$}} \\
KS continuous & \makecell[r]{$9.21$\\[-1.5pt]\scriptsize 0/5} & \makecell[r]{$14.77$\\[-1.5pt]\scriptsize 0/5} & \makecell[r]{$9.17$\\[-1.5pt]\scriptsize 0/5} & \makecell[r]{$14.65$\\[-1.5pt]\scriptsize 0/5} & \makecell[r]{$80.01$\\[-1.5pt]\scriptsize 0/5} & \makecell[r]{$1318.57$\\[-1.5pt]\scriptsize 1/5} & \makecell[r]{$7217.38$\\[-1.5pt]\scriptsize 1/5} \\
KS continuous, misspec. & \makecell[r]{$6.39$\\[-1.5pt]\scriptsize 0/5} & \makecell[r]{$16.90$\\[-1.5pt]\scriptsize 0/5} & \makecell[r]{$9.86$\\[-1.5pt]\scriptsize 0/5} & \makecell[r]{$14.93$\\[-1.5pt]\scriptsize 0/5} & \makecell[r]{$106.13$\\[-1.5pt]\scriptsize 0/5} & \makecell[r]{$4.00\mathrm{e}{+9}$\\[-1.5pt]\scriptsize 3/5} & \makecell[r]{$498.19$\\[-1.5pt]\scriptsize 0/5} \\
Beta-mixture continuous & \makecell[r]{$6.21$\\[-1.5pt]\scriptsize 0/5} & \makecell[r]{$17.81$\\[-1.5pt]\scriptsize 0/5} & \makecell[r]{$11.56$\\[-1.5pt]\scriptsize 0/5} & \makecell[r]{$14.83$\\[-1.5pt]\scriptsize 0/5} & \makecell[r]{$80.68$\\[-1.5pt]\scriptsize 0/5} & \makecell[r]{$1.66\mathrm{e}{+5}$\\[-1.5pt]\scriptsize 2/5} & \makecell[r]{$2143.92$\\[-1.5pt]\scriptsize 2/5} \\
SyntheticVCNet & \makecell[r]{$19.93$\\[-1.5pt]\scriptsize 0/5} & \makecell[r]{$21.84$\\[-1.5pt]\scriptsize 0/5} & \makecell[r]{$10.52$\\[-1.5pt]\scriptsize 0/5} & \makecell[r]{$18.02$\\[-1.5pt]\scriptsize 0/5} & \makecell[r]{$8.40$\\[-1.5pt]\scriptsize 0/5} & \makecell[r]{$288.56$\\[-1.5pt]\scriptsize 0/5} & \makecell[r]{$3.62\mathrm{e}{+4}$\\[-1.5pt]\scriptsize 3/5} \\
\midrule
\multicolumn{8}{l}{\textit{Synthetic, $N=2{,}000$}} \\
KS continuous & \makecell[r]{$20.06$\\[-1.5pt]\scriptsize 0/5} & \makecell[r]{$21.41$\\[-1.5pt]\scriptsize 0/5} & \makecell[r]{$20.49$\\[-1.5pt]\scriptsize 0/5} & \makecell[r]{$22.44$\\[-1.5pt]\scriptsize 0/5} & \makecell[r]{$107.15$\\[-1.5pt]\scriptsize 0/5} & \makecell[r]{$776.38$\\[-1.5pt]\scriptsize 0/5} & \makecell[r]{$1747.91$\\[-1.5pt]\scriptsize 1/5} \\
KS continuous, misspec. & \makecell[r]{$18.33$\\[-1.5pt]\scriptsize 0/5} & \makecell[r]{$22.82$\\[-1.5pt]\scriptsize 0/5} & \makecell[r]{$18.51$\\[-1.5pt]\scriptsize 0/5} & \makecell[r]{$33.13$\\[-1.5pt]\scriptsize 0/5} & \makecell[r]{$151.29$\\[-1.5pt]\scriptsize 0/5} & \makecell[r]{$6.06\mathrm{e}{+6}$\\[-1.5pt]\scriptsize 4/5} & \makecell[r]{$251.96$\\[-1.5pt]\scriptsize 0/5} \\
Beta-mixture continuous & \makecell[r]{$9.11$\\[-1.5pt]\scriptsize 0/5} & \makecell[r]{$33.73$\\[-1.5pt]\scriptsize 0/5} & \makecell[r]{$8.60$\\[-1.5pt]\scriptsize 0/5} & \makecell[r]{$19.46$\\[-1.5pt]\scriptsize 0/5} & \makecell[r]{$62.55$\\[-1.5pt]\scriptsize 0/5} & \makecell[r]{$3.32\mathrm{e}{+5}$\\[-1.5pt]\scriptsize 3/5} & \makecell[r]{$3844.31$\\[-1.5pt]\scriptsize 1/5} \\
SyntheticVCNet & \makecell[r]{$59.08$\\[-1.5pt]\scriptsize 0/5} & \makecell[r]{$34.25$\\[-1.5pt]\scriptsize 0/5} & \makecell[r]{$29.20$\\[-1.5pt]\scriptsize 0/5} & \makecell[r]{$38.54$\\[-1.5pt]\scriptsize 0/5} & \makecell[r]{$29.49$\\[-1.5pt]\scriptsize 0/5} & \makecell[r]{$217.80$\\[-1.5pt]\scriptsize 0/5} & \makecell[r]{$9058.52$\\[-1.5pt]\scriptsize 5/5} \\
\midrule
\multicolumn{8}{l}{\textit{Semi-synthetic}} \\
Digits & \makecell[r]{$2.65$\\[-1.5pt]\scriptsize 0/5} & \makecell[r]{$3.15$\\[-1.5pt]\scriptsize 0/5} & \makecell[r]{$2.09$\\[-1.5pt]\scriptsize 0/5} & \makecell[r]{$3.50$\\[-1.5pt]\scriptsize 0/5} & \makecell[r]{$3.13$\\[-1.5pt]\scriptsize 0/5} & \makecell[r]{$4.65\mathrm{e}{+8}$\\[-1.5pt]\scriptsize 3/5} & \makecell[r]{$214.65$\\[-1.5pt]\scriptsize 0/5} \\
ACIC & \makecell[r]{$2.76$\\[-1.5pt]\scriptsize 0/5} & \makecell[r]{$3.47$\\[-1.5pt]\scriptsize 0/5} & \makecell[r]{$2.13$\\[-1.5pt]\scriptsize 0/5} & \makecell[r]{$3.52$\\[-1.5pt]\scriptsize 0/5} & \makecell[r]{$7.76$\\[-1.5pt]\scriptsize 0/5} & \makecell[r]{$70.27$\\[-1.5pt]\scriptsize 0/5} & \makecell[r]{$5.48\mathrm{e}{+8}$\\[-1.5pt]\scriptsize 2/5} \\
California & \makecell[r]{$27.11$\\[-1.5pt]\scriptsize 0/5} & \makecell[r]{$12.58$\\[-1.5pt]\scriptsize 0/5} & \makecell[r]{$22.86$\\[-1.5pt]\scriptsize 0/5} & \makecell[r]{$12.54$\\[-1.5pt]\scriptsize 0/5} & \makecell[r]{$105.55$\\[-1.5pt]\scriptsize 0/5} & \makecell[r]{$1238.72$\\[-1.5pt]\scriptsize 1/5} & \makecell[r]{$49.77$\\[-1.5pt]\scriptsize 0/5} \\
\bottomrule
\end{tabular}
\end{adjustbox}
\caption{Maximum held-out weights of the selected weight estimators, by
continuous-treatment dataset and method. Each cell shows the largest
weight across all evaluation observations and five seeds/folds, with the number
of runs whose largest weight exceeds $10^3$ underneath.
The panels cover four synthetic datasets at each training size and three
semi-synthetic datasets, totaling 55 runs per method.
Weights are capped at $10^{12}$.}
\label{tab:causal_maximum_weights}
\end{table}

\subsection{Computation}\label{app:computation}

\begin{figure}[htbp]
  \centering
  \includegraphics[width=0.85\linewidth]{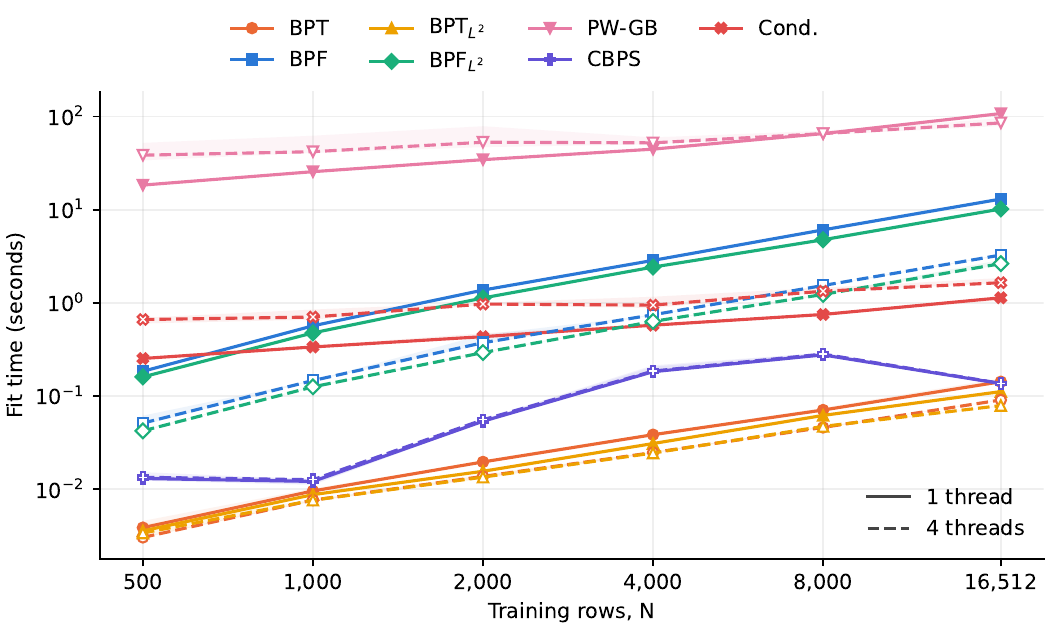}
  \caption{Fit-only wall-clock time on nested California subsamples with
  fixed selected configurations, on log--log axes. Solid lines use one thread;
  dashed lines allow up to four threads through each implementation's
  existing parallelism. Points show the median
  of three repetitions; shading shows their min--max range.
  CBPS uses its exactly identified
  fallback for some fits.}
  \label{fig:fit_time_vs_n}
\end{figure}

We measure fit-only scaling on California ($d=8$), fixing each
method's selected configuration from the first training fold (seed 0).
We fit each method on six nested subsamples of 500 to 16,512 rows, sharing the same
shuffled rows and generated treatments across methods.
With an untimed warmup per method and process, we time three fresh fits
per size in randomized order, running fits sequentially in separate processes
with one- and four-thread budgets on an Apple M4 Pro.
Both budgets use the same configurations and training data.
Timing excludes data preparation, cloning, tuning and prediction.
\Cref{fig:fit_time_vs_n} shows the resulting fit times.
\Cref{tab:causal_refit_times} reports refit times of the selected
configurations aggregated over all benchmark datasets and seeds/folds.
\Cref{tab:causal_fit_times} reports the full density-fitting time, which
adds the hyperparameter search to the refit.

\begin{table}[htbp]
\centering
\footnotesize
\setlength{\tabcolsep}{4pt}
\begin{adjustbox}{max width=\linewidth}
\begin{tabular}{lrrr}
\toprule
Method & Synthetic, $N=500$ & Synthetic, $N=2{,}000$ & Semi-synthetic \\
\midrule
BPT & \makecell[r]{${0.006}^{(2)}$\\[-1.5pt]$\pm 0.006$} & \makecell[r]{${0.018}^{(2)}$\\[-1.5pt]$\pm 0.013$} & \makecell[r]{${0.053}^{(2)}$\\[-1.5pt]$\pm 0.037$} \\
BPF & \makecell[r]{${0.040}^{(5)}$\\[-1.5pt]$\pm 0.027$} & \makecell[r]{${0.192}^{(5)}$\\[-1.5pt]$\pm 0.155$} & \makecell[r]{${0.932}^{(5)}$\\[-1.5pt]$\pm 0.654$} \\
BPT$_{L^2}$ & \makecell[r]{${\mathbf{0.006}}^{(1)}$\\[-1.5pt]$\pm 0.004$} & \makecell[r]{${\mathbf{0.017}}^{(1)}$\\[-1.5pt]$\pm 0.012$} & \makecell[r]{${\mathbf{0.051}}^{(1)}$\\[-1.5pt]$\pm 0.038$} \\
BPF$_{L^2}$ & \makecell[r]{${0.034}^{(4)}$\\[-1.5pt]$\pm 0.023$} & \makecell[r]{${0.157}^{(4)}$\\[-1.5pt]$\pm 0.123$} & \makecell[r]{${0.763}^{(4)}$\\[-1.5pt]$\pm 0.470$} \\
PW-GB & \makecell[r]{${27.901}^{(7)}$\\[-1.5pt]$\pm 29.453$} & \makecell[r]{${62.085}^{(7)}$\\[-1.5pt]$\pm 77.239$} & \makecell[r]{${59.364}^{(7)}$\\[-1.5pt]$\pm 67.147$} \\
CBPS & \makecell[r]{${0.029}^{(3)}$\\[-1.5pt]$\pm 0.009$} & \makecell[r]{${0.067}^{(3)}$\\[-1.5pt]$\pm 0.023$} & \makecell[r]{${0.174}^{(3)}$\\[-1.5pt]$\pm 0.146$} \\
Cond. & \makecell[r]{${2.006}^{(6)}$\\[-1.5pt]$\pm 2.071$} & \makecell[r]{${2.414}^{(6)}$\\[-1.5pt]$\pm 1.452$} & \makecell[r]{${5.037}^{(6)}$\\[-1.5pt]$\pm 1.145$} \\
\bottomrule
\end{tabular}
\end{adjustbox}
\caption{Selected-model fit-only
wall-clock time in seconds: each tuned configuration is refitted once on its
training sample, excluding search, data generation and prediction. Refits run
sequentially. Entries are mean $\pm$ sample SD over all datasets and
seeds/folds of the panel; the SD combines dataset and seed variation.
Parenthesized superscripts rank unrounded column means, with midranks for ties
and bold for the smallest mean. CBPS is untuned.}
\label{tab:causal_refit_times}
\end{table}

\begin{table}[htbp]
\centering
\footnotesize
\setlength{\tabcolsep}{4pt}
\begin{adjustbox}{max width=\linewidth}
\begin{tabular}{lrrr}
\toprule
Method & Synthetic, $N=500$ & Synthetic, $N=2{,}000$ & Semi-synthetic \\
\midrule
BPT & \makecell[r]{${0.403}^{(2)}$\\[-1.5pt]$\pm 0.102$} & \makecell[r]{${1.298}^{(3)}$\\[-1.5pt]$\pm 0.646$} & \makecell[r]{${6.510}^{(3)}$\\[-1.5pt]$\pm 3.757$} \\
BPF & \makecell[r]{${1.668}^{(5)}$\\[-1.5pt]$\pm 0.744$} & \makecell[r]{${8.049}^{(5)}$\\[-1.5pt]$\pm 5.173$} & \makecell[r]{${63.778}^{(5)}$\\[-1.5pt]$\pm 49.912$} \\
BPT$_{L^2}$ & \makecell[r]{${0.412}^{(3)}$\\[-1.5pt]$\pm 0.116$} & \makecell[r]{${1.232}^{(2)}$\\[-1.5pt]$\pm 0.598$} & \makecell[r]{${5.881}^{(2)}$\\[-1.5pt]$\pm 3.256$} \\
BPF$_{L^2}$ & \makecell[r]{${1.458}^{(4)}$\\[-1.5pt]$\pm 0.674$} & \makecell[r]{${6.789}^{(4)}$\\[-1.5pt]$\pm 4.296$} & \makecell[r]{${58.998}^{(4)}$\\[-1.5pt]$\pm 45.598$} \\
PW-GB & \makecell[r]{${67.580}^{(7)}$\\[-1.5pt]$\pm 53.225$} & \makecell[r]{${97.929}^{(6)}$\\[-1.5pt]$\pm 73.816$} & \makecell[r]{${145.672}^{(6)}$\\[-1.5pt]$\pm 99.028$} \\
CBPS & \makecell[r]{${\mathbf{0.037}}^{(1)}$\\[-1.5pt]$\pm 0.014$} & \makecell[r]{${\mathbf{0.070}}^{(1)}$\\[-1.5pt]$\pm 0.024$} & \makecell[r]{${\mathbf{0.292}}^{(1)}$\\[-1.5pt]$\pm 0.229$} \\
Cond. & \makecell[r]{${36.641}^{(6)}$\\[-1.5pt]$\pm 16.777$} & \makecell[r]{${195.080}^{(7)}$\\[-1.5pt]$\pm 120.260$} & \makecell[r]{${313.435}^{(7)}$\\[-1.5pt]$\pm 131.094$} \\
\bottomrule
\end{tabular}
\end{adjustbox}
\caption{Weight-estimation
wall-clock time in seconds, including the 25-trial permutation-log-loss search
and the selected-model refit, excluding data generation and prediction. Entries are mean $\pm$ sample SD over all datasets and
seeds/folds of the panel; the SD combines dataset and seed variation.
Parenthesized superscripts rank unrounded column means, with midranks for ties
and bold for the smallest mean. CBPS is untuned. The searches ran at different times under different
machine loads.}
\label{tab:causal_fit_times}
\end{table}

\subsection{Dose-response curves}\label{app:dose-response}

\begin{figure}[htbp]
  \centering
  \includegraphics[width=\linewidth]{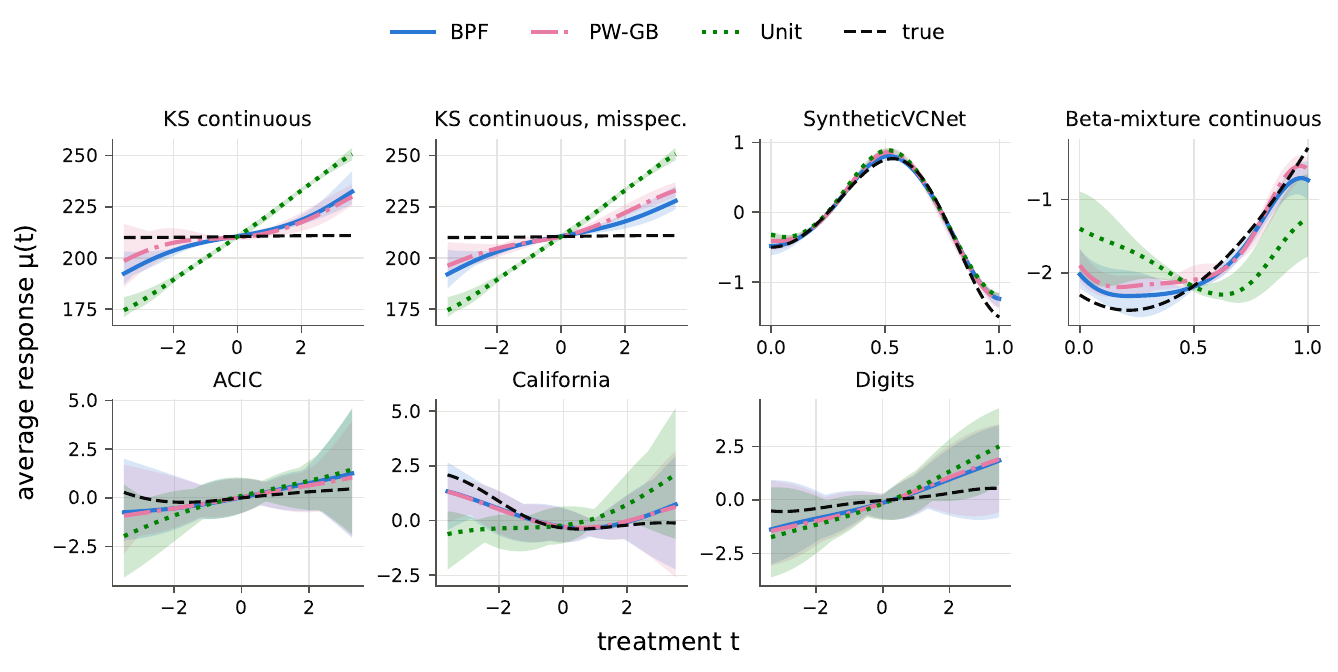}
  \caption{MSM response curves of BPF, PW-GB and Unit, the constant weight $\widehat w\equiv1$.
  Top: the four continuous synthetic settings at $N=2{,}000$.
  Bottom: the three semi-synthetic settings. Lines show mean estimates over five seeds/folds, and shaded bands show the
  min--max range across them; the dashed black line
  is the true response $\mu(t)$. Each semi-synthetic fold has its own response curve, so in these settings the
  dashed line shows the average of five true curves and the bands mostly reflect
  variation across those curves.}
  \label{fig:causal_response_examples}
\end{figure}

\begin{figure}[htbp]
  \centering
  \includegraphics[width=\linewidth]{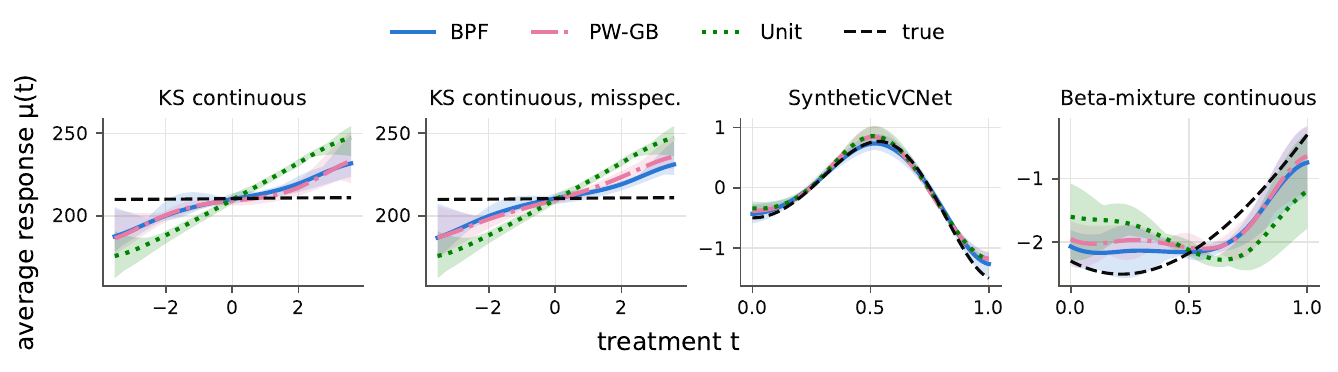}
  \caption{MSM dose-response curves, continuous synthetic settings, $N=500$. Lines show mean estimates over five seeds/folds, and shaded bands show the
  min--max range across them; the dashed black line
  is the true response $\mu(t)$.}
  \label{fig:dose_response_500}
\end{figure}

\begin{figure}[htbp]
  \centering
  \includegraphics[width=\linewidth]{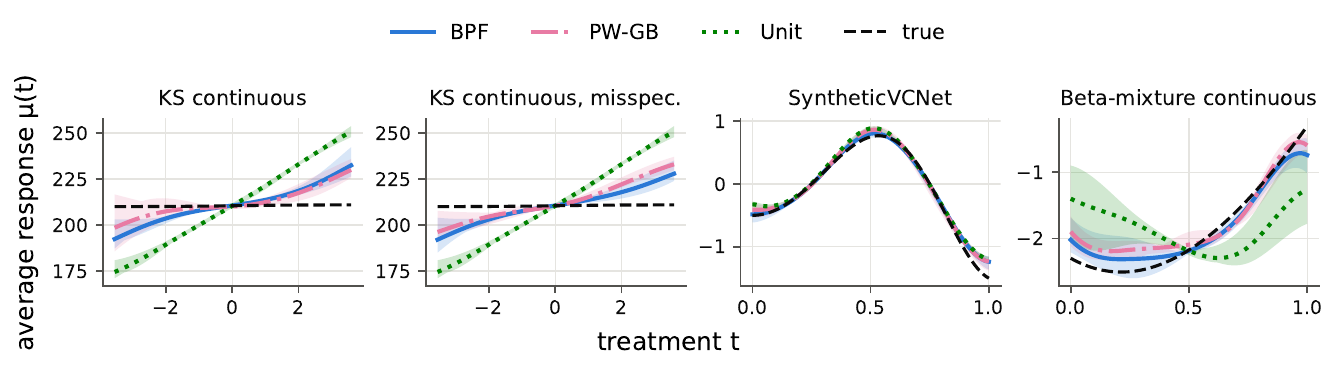}
  \caption{MSM dose-response curves, continuous synthetic settings, $N=2{,}000$. Lines show mean estimates over five seeds/folds, and shaded bands show the
  min--max range across them; the dashed black line
  is the true response $\mu(t)$.}
  \label{fig:dose_response_2000}
\end{figure}

\begin{figure}[htbp]
  \centering
  \includegraphics[width=\linewidth]{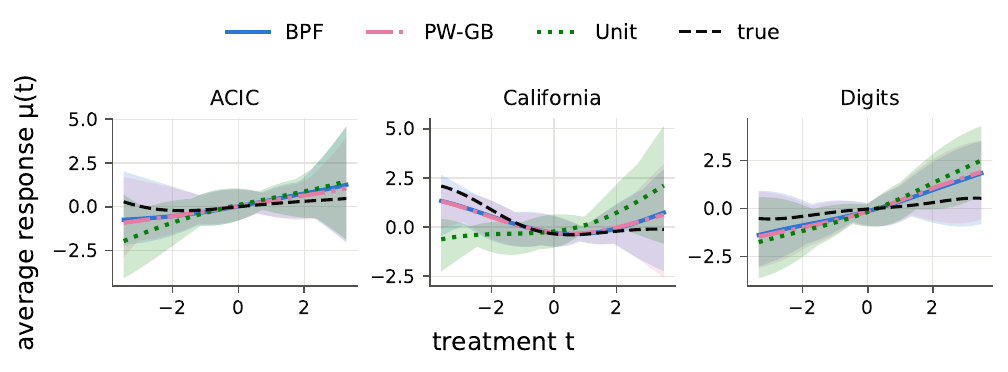}
  \caption{MSM dose-response curves, semi-synthetic settings. Lines show mean estimates over five seeds/folds, and shaded bands show the
  min--max range across them; the dashed black line
  is the true response $\mu(t)$. Each semi-synthetic fold has its own response curve, so in these settings the
  dashed line shows the average of five true curves and the bands mostly reflect
  variation across those curves.}
  \label{fig:dose_response_semi}
\end{figure}

\Cref{fig:causal_response_examples,fig:dose_response_500,fig:dose_response_2000,fig:dose_response_semi}
show the MSM response curves behind the RMSEs of
\cref{tab:causal_msm_response}.
\Cref{fig:causal_response_examples} combines the $N=2{,}000$ synthetic and the semi-synthetic settings in one figure.
\Cref{fig:dose_response_500,fig:dose_response_2000} cover the continuous
synthetic settings at $N=500$ and $N=2{,}000$, and
\cref{fig:dose_response_semi} covers the semi-synthetic settings. Besides BPF and PW-GB, each panel includes
\textbf{Unit}, the constant weight $\widehat w \equiv 1$ that yields the
unweighted estimator. The gap between its estimated curve and the true curve shows how much confounding
the weights have to remove.

\end{document}